\documentclass[aps,prb,preprint,groupedaddress]{revtex4-2}

\usepackage{graphicx}
\usepackage{dcolumn}
\usepackage{bm}
\usepackage{xurl}
\usepackage{hyperref}
\usepackage[mathlines]{lineno}
\usepackage{bibunits}
\usepackage{geometry}
\usepackage{setspace}
\usepackage{etoolbox}
\apptocmd{\thebibliography}{\raggedright}{}{}
\usepackage{booktabs}
\usepackage{amsmath}

\AtBeginDocument{%
  \setlength{\emergencystretch}{3em}
  \sloppy
}

\def\vte{VTe\textsubscript{2}} 
\def\vse{VSe\textsubscript{2}} 
\def\vs{VS\textsubscript{2}} 
\def\hvte{H-VTe\textsubscript{2}}

\def\vx{V\textit{X}\textsubscript{2}} 
\def\hvx{H-V\textit{X}\textsubscript{2}}
\def\hvse{H-VSe\textsubscript{2}}
\def\hvs{H-VS\textsubscript{2}}

\usepackage[final]{changes}  

\begin{document}

\begin{bibunit}[apsrev4-2]

\title{
Transferable mechanism  of perpendicular magnetic anisotropy switching by hole doping in VX\textsubscript{2} (X=Te, Se, S) monolayers
}

\author{John Lawrence Euste}

\affiliation{Scuola Internazionale Superiore di Studi Avanzati (SISSA), Trieste, Italy}
\affiliation{The Abdus Salam International Centre for Theoretical Physics (ICTP), Trieste, Italy}

\author{Maha Hsouna}

\affiliation{Scuola Internazionale Superiore di Studi Avanzati (SISSA), Trieste, Italy}
\affiliation{The Abdus Salam International Centre for Theoretical Physics (ICTP), Trieste, Italy}

\author{Nata\v sa Stoji\' c}

\affiliation{The Abdus Salam International Centre for Theoretical Physics (ICTP), Trieste, Italy}
\email[]{nstojic@ictp.it}

\begin{abstract}
The ability to tune and switch magnetic anisotropy to a perpendicular orientation is a key challenge for implementing two-dimensional magnets in spintronic devices.
H-phase vanadium dichalcogenides \vx{} ($X$=Te, Se, S) are promising ferromagnetic semiconductors  with large magnetic anisotropy energy (MAE). Recent work has shown that hole doping can switch their easy axis to out-of-plane, although the microscopic origin of this perpendicular magnetic anisotropy (PMA) remains unclear.
 Using density-functional-theory calculations, we demonstrate that the PMA enhancement arises from first-order spin-orbit coupling (SOC) acting on topmost degenerate valence states with nonzero orbital angular momentum projection ($m_l\ne 0$).  In this case,   the $\hat{L}_z \hat{S}_z$ term dominates for perpendicular magnetization orientation,  while in-plane orientations involve only weaker, second-order SOC contributions.
 The increased valence bandwidth leads to depletion of higher-energy states upon hole doping, stabilizing PMA. From this mechanism,
we identify two transferable design principles for enhancing  magnetic anisotropy under weak hole doping: (i) orbital degeneracy at the valence-band edge protected by point-group symmetry   and (ii) finite SOC in the degenerate manifold.
Notably, we identify multiple magnetic semiconductors that meet these criteria and display enhanced MAE under hole doping. Furthermore, we show that band engineering can strategically place these degenerate orbitals at the valence band edge, significantly boosting PMA when hole-doped. 
We also examine trends in \vte{}, \vse{}, and \vs{} to determine the influence of crystal-field splitting, exchange interaction, and orbital hybridization on the valence band edges. 
These results provide both a fundamental understanding of PMA switching upon hole doping and a transferable strategy for tuning magnetic anisotropy, essential for designing high-performance spintronic materials.

\end{abstract}

\maketitle


\section{Introduction}

Two-dimensional (2D) materials with magnetization perpendicular to the surface can be utilized to develop efficient, robust and compact spintronic devices\cite{hu2018engineering, naganuma2020perpendicular}.
Various strategies have been developed to control the magnetic anisotropy energy (MAE) and manipulate the easy axis of magnetization, including strain engineering \cite{tang2022strain, ZhaLiuXu23, chen2024recent}, stacking \cite{zhang2019ferromagnet,ruiz2023stacking}, and doping \cite{hu2018engineering,YadStoBin24}.
Electrostatic doping, for example, can increase the MAE and switch the easy axis of magnetization in some monolayers \cite{kim2019exploitable, guo2024significant, saritas2022piezoelectric, cheng2021large, an2024carrier, han2023hole, ma2020intrinsic, ren2022manipulating, doi:10.1021/acs.nanolett.9b03316, kim2021drastic, yao2023control, chen2020electronic, sheng2021magnetic, wang2021effects, chen2024first,  jiang2024electronic} due to the changes in the electronic structure near the Fermi level caused by emptying/occupying bands upon hole/electron doping. Techniques such as applying electric gate voltage and exposing the surface to certain molecular reductants or oxidants  demonstrate that charge doping is feasible in experiments \cite{zhang2018controllable, wang2023controllable}.
\\

The search for perpendicular magnetic anisotropy (PMA) has been largely focused on 2D transition metal dichalcogenides (TMD) that exhibit ferromagnetic ordering with high Curie temperature and possess weak interlayer van der Waals interactions, making them ideal for low-dimensional structures. The successful experimental synthesis \cite{feng2011metallic} of \vs{} drew attention to vanadium dichalcogenides \vx{} ($X=$ S, Se, or Te) for their versatile band structure and tunable magnetic properties.
The ground state of the \vs{} monolayer is a ferromagnetic semiconductor. Its monolayer crystallizes in a trigonal prismatic structure, also known as the H-phase.
Similar electronic and magnetic properties have been observed in \hvse{} and \hvte{} \cite{fuh2016newtype, abdul2022electronic, jafari2023electronic}.
Vanadium dichalcogenides have recently gained significant attention due to their intriguing physical properties and broad potential applications in photocatalysis, optoelectronics, and spintronics \cite{abdul2022electronic,jafari2023electronic,tariq2021pristine,liu2023magnetic}. \\

Density functional theory (DFT) calculations revealed that \hvx{} monolayers all have in-plane (IP) easy magnetization axis which can be switched to the out-of-plane (OOP) direction by hole doping \cite{chen2020electronic, sheng2021magnetic, chen2024first, jiang2024electronic,zhang2025effect}. Yet, the mechanism behind hole-doping-induced PMA in \hvx{} and related cases remains unclear despite the comprehensive understanding of the easy-plane magnetization in the pristine cases usually based on interpretations from second-order perturbation theory \cite{chen2020electronic, sheng2021magnetic, tang2022strain, wang2021effects, jafari2023electronic}. \\

Most studies of magnetic anisotropy employ force theorem \cite{wang1996validity} and second-order perturbation theory \cite{wang1993first} in their analysis.
However, both the force theorem and the second-order perturbation theory lose their validity for systems with strong spin-orbit coupling (SOC) \cite{blanco2019validity}. In addition, the force theorem has been reported to be less reliable in low-dimensional structures. \cite{BloLehDen10} 
At the same time, the second order perturbation theory approach, which can be expressed through virtual transitions between occupied and empty states, has difficulties dealing with the addition or removal of bands resulting from the shift of the Fermi level upon even the slightest doping, possibly causing an infinite MAE \cite{smiri2021dft+}.
Such changes lie beyond perturbative treatment and could only be addressed for minimal doping levels through ad hoc redefinition of valence and conduction bands---a case-specific approach not supported by standard electronic structure codes.
Another custom-tailored approach used to interpret the changes of MAE upon hole doping adopts second-order perturbation theory to compute energy-resolved contributions to the MAE and approximates the effect of doping by selectively adding or removing these energy-dependent  terms from the pristine MAE \cite{saritas2022piezoelectric}. This approach is applicable only under very weak doping and within the rigid band approximation. 
\\

A clear and intuitive understanding of the hole-doping-induced transition to PMA based on the knowledge of the highest valence states (and not dependent, for example, on the matrix elements between occupied and unoccupied states), along with the essential conditions for such transition applicable in a range of systems, could provide predictive framework that would enable targeted material discovery and strategic band engineering to realize such conditions. Furthermore, identification and design of materials with tunable MAE upon hole doping would be significantly advanced by a deeper mechanistic insight into how the interplay of crystal field, exchange splitting, and orbital hybridization influences the highest valence states.
\\

In this work, we employ DFT to evaluate and analyze the trends in MAE of H-phase vanadium dichalcogenide (\hvx{}) monolayers, aiming to elucidate the mechanism governing magnetic anisotropy changes under charge doping, with particular focus on the weak hole-doping regime.
We investigate trends from $X=$ Te to Se to S in crystal field splitting, exchange interaction, and orbital hybridization, which influence the topmost valence bands.
We show here that the switching to PMA in the hole-doped systems can be attributed to the larger spin-orbit splitting effect of the valence band maximum when spins are aligned OOP, relative to the IP orientation.
Building on this understanding, we establish a set of straightforward and simple conditions within the scalar-relativistic framework to enhance MAE in magnetic semiconductors under hole doping, governed by orbital degeneracies of the valence band maximum and finite atomic SOC.
We confirm these conditions by identifying multiple material examples from the literature that exhibit MAE enhancement under hole doping, all of which satisfy our proposed criteria.
Using ab initio DFT+U calculations, we show that the doping-induced trends in MAE and the mechanism of PMA switching are not significantly affected by the variations in the effective Hubbard $U$ parameter for pristine \hvx{}.
Furthermore, we analyze the trends in magnetic moment with hole doping based on the spin polarization of the emptied bands. Finally, guided by our established design criteria, we engineer the band structure of \hvs{} to significantly enhance the MAE and achieve PMA switching at reduced hole doping concentrations.
\\

Our findings are presented as follows: After the method section, we describe the structure and some properties of the \hvx{} monolayers in the section "System".
In the results section, we first focus on \hvte{}, the strongest-SOC member among \vx{} compounds, \added{possessing only one VBM}.
We discuss the effect of charge doping on its MAE and explain the mechanism behind its switching to PMA upon hole doping in this case. 
Subsequently, we \added{apply this mechanism to explain the MAE switching in \vse{} and \vs{}, which are characterized by weaker SOC and by the presence of two VBMs close in energy. We then}
examine the trends \added{in MAE} across \vte{}, \vse{}, and \vs{} to identify how crystal field splitting, exchange interaction, and orbital hybridization influence  the valence band edges and, consequently, the onset of PMA upon hole doping.
\deleted{Furthermore, we present and clarify the trends in the V magnetic moment in hole-doped \hvx{}. In the next subsection, we discuss the effect of Hubbard $U$ parameter on MCA and switching to PMA.}
Finally, we show how the presented mechanism can be applied to other related materials and used to band engineer electronic structure to obtain faster onset and enhanced strength of PMA through hole doping.

\section{Methods}
The structural, electronic, and magnetic properties of vanadium dichalcogenides V$X_2$ ($X=$ Te, Se, or S) in the H-phase were calculated using density functional theory (DFT) as implemented in Quantum Espresso \cite{qe1}. We employed the generalized gradient approximation (GGA) with the Perdew-Burke-Ernzerhof (PBE) functional, using both scalar and fully relativistic projector-augmented wave (PAW) pseudopotentials.
As the electrons in the V $3d$ orbitals are strongly correlated, we employed the GGA+$U$ approach with an effective Hubbard $U$ of 1.3~eV.
\added{This value reproduces key electronic properties—such as the band gap and the position of the VBM—that are relevant for our MAE analysis and are consistent with previous studies, as shown in the Results section. }
We present the results with this value of $U$ for each \vx{} monolayer, but, as we show in this work, our findings about MAE trends can be extended to calculations at the level of $U=0$ and 1.7~eV which fall within the range of $U$ values adopted in the literature \cite{sheng2021magnetic,jafari2023electronic,chen2020electronic, wang2021effects, tang2022strain, zhang2025effect, jiang2024electronic,chen2024first}.

For the scalar-relativistic calculations, a plane-wave basis with an energy cutoff of at least 60~Ry for the wavefunctions and 640~Ry for the density was used. The Brillouin zone (BZ) was sampled using a $\Gamma$-centered Monkhorst-Pack $12\times 12\times 1$ grid of k-points.
A supercell approach was employed where a vacuum spacing of 20~Å along $\hat{z}$ was introduced to avoid interlayer interactions.
Equilibrium configurations for the FM and AFM structures were obtained by relaxing the atomic coordinates with a force convergence threshold of $10^{-3}$~Ry/bohr.
Spin-orbit coupling effects were included by using fully relativistic pseudopotentials with noncollinear calculations to allow both IP and OOP magnetization. The sampling grid in the BZ was increased to $26\times 26\times 1$ for calculations with SOC. The computational parameters were chosen for each \vx{} system so that the total energy differences are within an error of less than 0.03~meV \added{- see Supplementary Section~S1 for more details \cite{supp}.}

\nocite{murila2021structural,su2021recent,joseph2023hydrothermally}

Phonon frequencies were calculated using density functional perturbation theory to determine structural stability. The dynamical matrices were calculated on a $4\times 4 \times 1$ q-point grid with a $10^{-16}$~Ry convergence threshold.

The magnetocrystalline anisotropy energy ($E_{MCA}$) was computed as the difference between the total energy of the structure with IP magnetization (along the surface plane, $E^\parallel$) and that with OOP magnetization (along the surface normal, $E^\perp$), both obtained from fully relativistic self-consistent calculations:

\begin{equation}
\label{eq:mca}
E_{MCA} = E^\parallel - E^\perp = E^x - E^z.
\end{equation}

$E_{MCA}$ is the main contribution to the MAE which can be expressed as the sum of $E_{MCA}$ and $E_{SA}$, where $E_{SA}$ stands for the shape anisotropy energy.
A negative/positive MAE value implies easy in-plane/out-of-plane magnetization.
The $E_{SA}$ contribution to the MAE is evaluated for the \vte{} monolayer by summing all unique pairwise dipolar interaction energies between magnetic moments, as in  Ref.~\cite{BroGhoDeb24}, using the DFT values for the \vte{} lattice parameter and V magnetic moment. Additional details about the calculation of $E_{SA}$ for all three systems, both doped and pristine, are given in Supplementary Section~S2.  The MCA and the V magnetic moments were calculated at different charge carrier doping levels $\delta$. Vacuum spacing was increased to 25~Å to avoid interlayer interactions after charge doping. Charge-doped systems were obtained by adding holes or electrons and using a compensating jellium background in the supercell to maintain charge neutrality. Relaxations were performed for each doping concentration with scalar-relativistic GGA+U. Band structures are referenced to the vacuum level \added{(which is the planar average of the electrostatic potential in the vacuum region separating periodic images of monolayers)}, enabling absolute comparisons of the VBM energies across different magnetization directions.

To probe the changes in MAE and magnetic moments from the underlying band structures, we looked into the spin and orbital characters of the bands, especially at the valence band edges. The spin $S_\alpha$ character ($\alpha=x$ for IP and $\alpha=z$ for OOP) of the bands in $k$-space is calculated as the expectation value of the spin operator  ${\frac{1}{2}}\hat{\sigma}_\alpha$ over the Bloch spinor eigenfunctions $\Psi_{n, { k}}({ r})$, given by
$ S_\alpha(n,{ k}) = \frac{1}{2}  {\langle\Psi_{n, { k}}|\hat{\sigma}_\alpha|\Psi_{n, {k}}\rangle } / {\langle\Psi_{n, { k}}|\Psi_{n, { k}}\rangle} $ where $n$ is the band index and $\hat{\sigma}_{\alpha}$ are the Pauli matrices. The contribution $c_{lm_l,k}$ of each atomic orbital $| \phi_{nlm_l} \rangle$ to a given wavefunction $| \Psi_{n,k} \rangle$ can be calculated by projecting the wavefunctions at a given $k$-point to each orbital:$| \Psi_{n,k} \rangle = \Sigma_{lm_l} c_{lm_l,k} | \phi_{nlm_l} \rangle$.

\section{System}

The H-phase monolayer of \vx{} consists of one vanadium plane positioned  between two chalcogen ($X=$ Te, S, or Se) planes such that each V atom in the unit cell is surrounded by six $X$ atoms in a trigonal prismatic coordination.
The H phase is the ground state for \vs{}, whereas \vte{} and \vse{} stabilize in the charge-density-wave T-phase as their lowest-energy configuration \cite{tang2022strain,abdul2022electronic} with many proposed mechanisms of transition to the H phase  \cite{tang2022strain, zhu2022charge, wang2021effects}. 

\begin{figure}[h]
    \centering
    \includegraphics[width=0.7\textwidth]{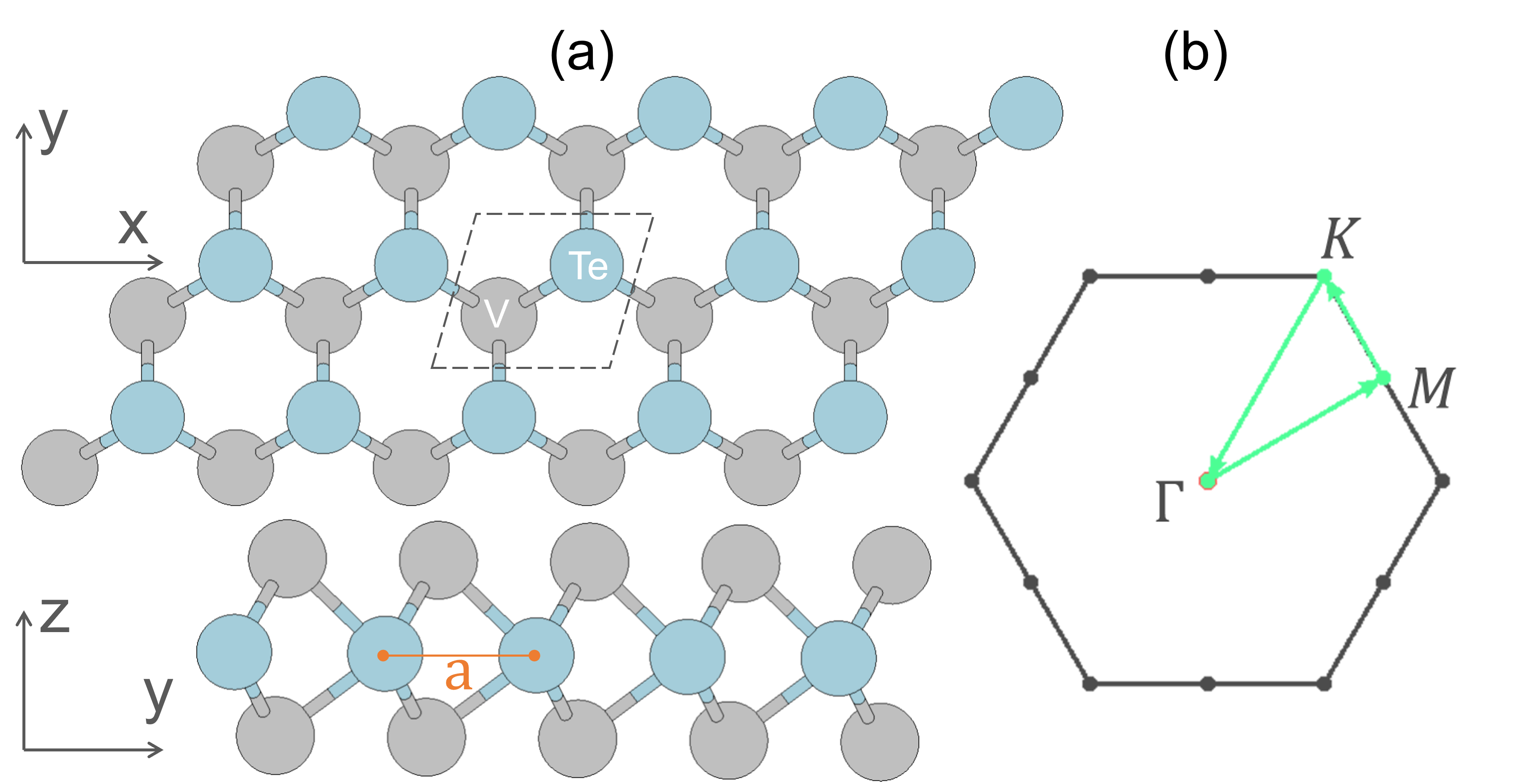}
    \caption{a) Trigonal prismatic structure of \hvte{} (blue: V, gray: Te) with the top view in the upper panel and side view in the lower panel. The parallelogram marks the unit cell with lattice constant $a$. Other vanadium dichalcogenide monolayers in the H phase have the same type of structure replacing Te with Se or S. b) \replaced{The path in the Brillouin zone used in the band structure calculations. }{Phonon dispersion of \hvte{}.  The inset illustrates the path in the Brillouin zone used in the phonon and band structure calculations.}} 
    \label{fig:hvte2_struct}
\end{figure}

Fig.~\ref{fig:hvte2_struct}a illustrates the structure of \hvte{}; other vanadium dichalcogenides have the same structural arrangement but different parameters when replacing Te with Se or S. \vte{} has the largest unit cell with the optimized lattice constant of $a=3.64$~\AA, followed by \vse{} with $a=3.33$~\AA, while the lattice constant of \vs{} is the smallest with $a=3.19$~\AA.
The system's crystallographic symmetry is described by the $D_{3h}$ point group combining a threefold rotation axis, three perpendicular twofold axes, a horizontal mirror plane, and three vertical mirror planes. The 12 symmetry operations define the system's trigonal prismatic structure. Crucially, in the scalar-relativistic case, $D_{3h}$ symmetry enforces degeneracies in the  $d_{xz}/d_{yz}$  and $d_{x^2-y^2}/d_{xy}$ orbital pairs at $\Gamma$, while at the K point both degeneracies are preserved thanks to the $C_{3h}$ small group at K \added{(see Fig.~\ref{fig:hvte2_struct}b for the high-symmetry points in the Brillouin zone)}.
Ferromagnetism with SOC breaks the $D_{3h}$ symmetry of \hvx{}, preserving only $C_{2v}$ (for IP orientation of spins). This corresponds to Shubnikov Type III, where spin alignment removes the 3-fold axis but retains two mirror planes. The magnetic space group in this case is P$2mm'$  and P$\bar{3}'m'1$ for the OOP spin direction.

We have confirmed that the ferromagnetic (FM) semiconducting H-\vte{} monolayer is dynamically stable, as evidenced by the absence of imaginary frequencies in its phonon spectrum: see SI Section S1. Likewise, within the H-phase structure,  \vse{} and \vs{} are ferromagnetic semiconductors. We therefore focus on the FM structures when calculating their magnetic anisotropy.

\section{Results and discussion}

\subsection{Changes in the magnetic anisotropy with charge doping in \hvte{}}

\begin{figure}[h]
    \centering
    \includegraphics[width=0.7\textwidth]{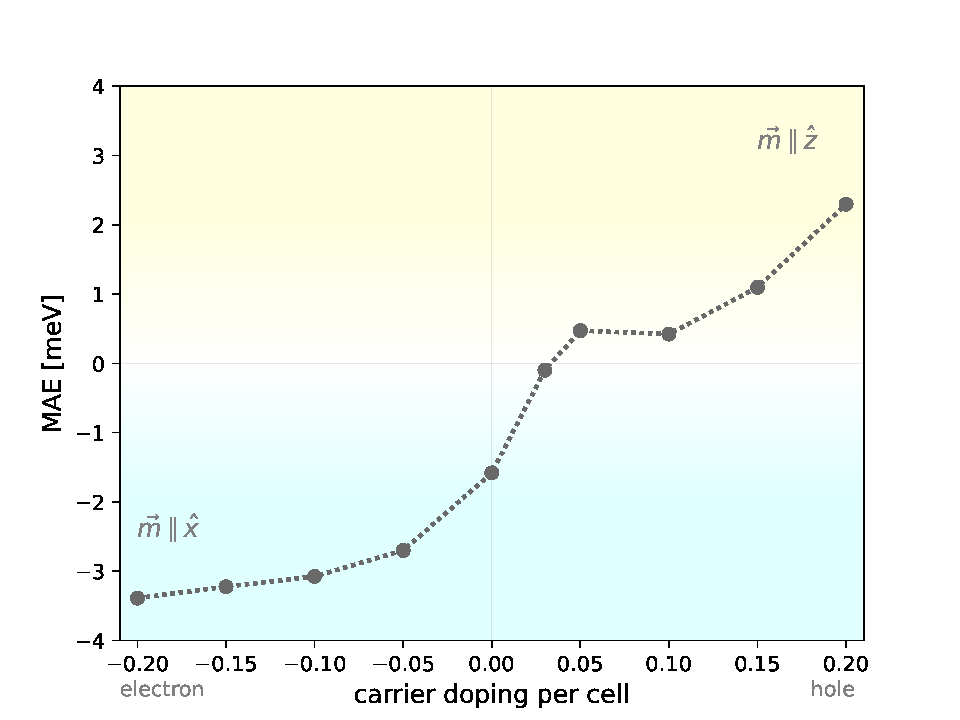}
    \caption{MAE versus charge carrier doping concentrations $\delta$ for FM \hvte{}. The calculated $E_{MCA}$ corresponds to the MAE since $E_{SA}$ is  15~$\mu$eV for the pristine case, and smaller than 18~$\mu$eV for all doping levels shown in the plot (see Supplementary Section~S3). Hole doping corresponds to $\delta>0$ while electron doping $\delta<0$.}
    \label{fig:mca_dope_U1.3}
\end{figure}

MCA of the undoped or pristine \hvte{} monolayer was first calculated using Eq. \ref{eq:mca} by setting the IP magnetization along the $x$-axis and the OOP magnetization along $z$. We obtain $E_{MCA}=-1.58$~meV indicating easy in-plane magnetization, in agreement with other studies \cite{jafari2023electronic,tang2022strain,chen2020electronic,zhang2025effect}.
Since our calculated shape anisotropy (see Supplementary Section~S2) is negligible ($E_{SA}=15~\mu\text{eV}$) relative to the MCA, we equate the MCA  to the MAE. 
To see how MAE changes with electron ($\delta<0$) and hole ($\delta>0$) doping per unit cell, MAE values at different amounts of doping are plotted in Fig.~\ref{fig:mca_dope_U1.3}.
Electron doping maintains the easy IP magnetization while hole doping shifts to stable OOP magnetization with MAE rising with increasing hole doping concentration.
The MAE switches to positive already for a small hole doping, e.g. $+0.47$~meV MAE at $\delta=+0.05$~h/cell, implying strong PMA in \hvte{} with hole doping. \added{Based on linear interpolation, we estimate the transition to PMA to occur at a hole doping of $\delta = +0.033$~h/cell.}

\subsection{Mechanism of PMA switching in hole-doped \hvte{}}

\begin{figure}[h!]
    \centering
    \includegraphics[width=1\textwidth]{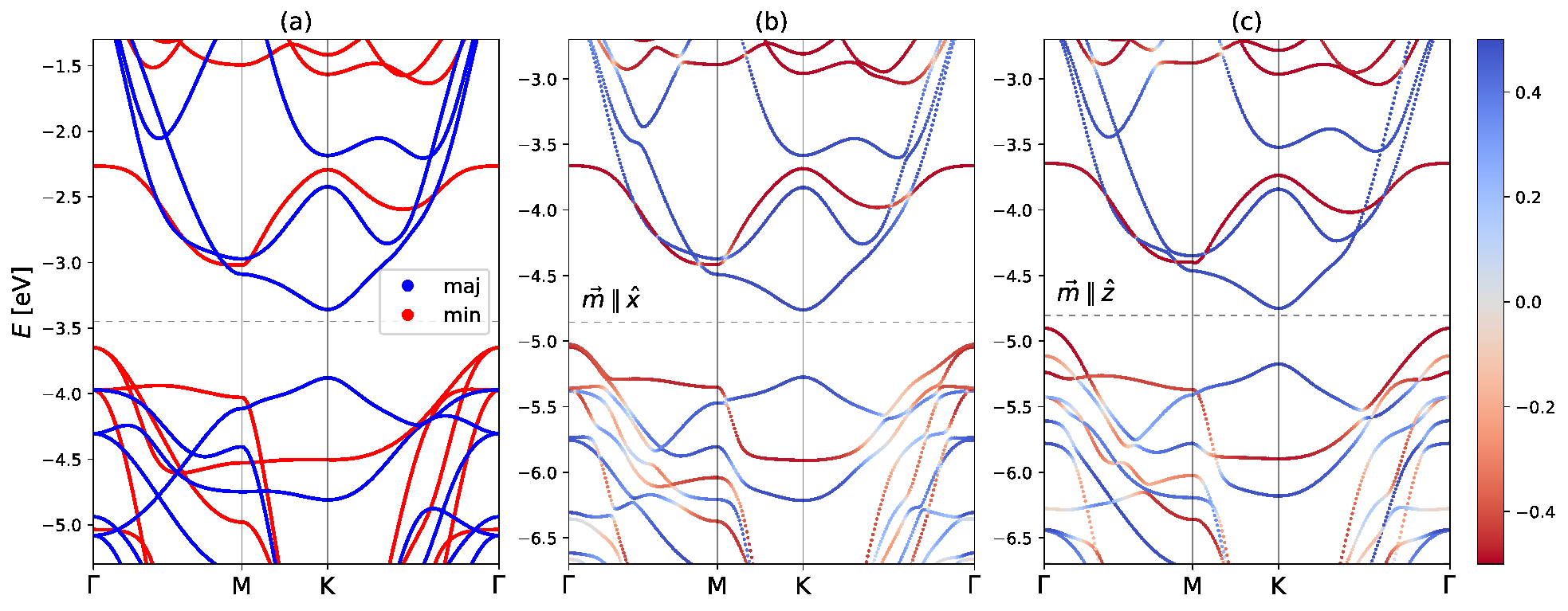}
    \caption{(a) Spin-polarized band structure of \hvte{}; the blue/red bands represent majority/minority spin channels. Fully relativistic spin-projected bands for pristine \hvte{} with (b) in-plane and (c) out-of-plane magnetization. The energies are referenced to the vacuum level and the dotted line denotes the Fermi level. The bar at the right indicates the color code for the band projection of the spins along $x$ in (b) and along $z$ in (c). }
    \label{fig:sp_bands_u1.3}
\end{figure}

The spin-polarized band structure for $\delta=0$ in Fig.~\ref{fig:sp_bands_u1.3}a confirms that \hvte{} is an indirect band gap ($E_g=0.29$~eV) semiconductor with the conduction band minimum (CBM) at the K point and the valence band maximum (VBM) at $\Gamma$, in good agreement with previous calculations \cite{fuh2016newtype, jafari2023electronic, tang2022strain}.
The states near the Fermi level exhibit a dominant vanadium d-orbital character,  as confirmed by the orbital-projected bands shown in Supplementary Section~S3.
The VBM at $\Gamma$ has minority-spin electrons in V $d_{xz}$ and $d_{yz}$ orbitals hybridized with Te $p_x$ and $p_y$, respectively.
The presence of this Te $p$ character at the VBM secures sizable spin-orbit splitting. 
In Fig.~\ref{fig:sp_bands_u1.3}b and \ref{fig:sp_bands_u1.3}c we show the fully relativistic band structures for the magnetization axis along $x$ and along $z$.
The key difference between the IP and OOP band structures occurs at the $\Gamma$-point VBM: the OOP splitting (337~meV) is significantly larger than the IP splitting (18~meV), shifting the VBM to higher energy in the OOP case.
The energy gap is different for the two magnetization directions: 0.29~eV for IP and 0.15~eV for OOP.

We explain the effects of electrostatic doping on MAE and magnetic moment primarily from the band structures.  Electrostatic doping shifts the Fermi level $E_F$ due to the partial occupation/emptying of the conduction/valence bands. The valence and conduction band edges near $E_F$ are each dominated by a single spin channel, i.e. majority spin in the conduction band edge and minority in the valence, so weak electron or hole doping renders \hvte{} a half-metal.

\begin{figure}[h!]
    \centering    
    \includegraphics[width=1\linewidth]{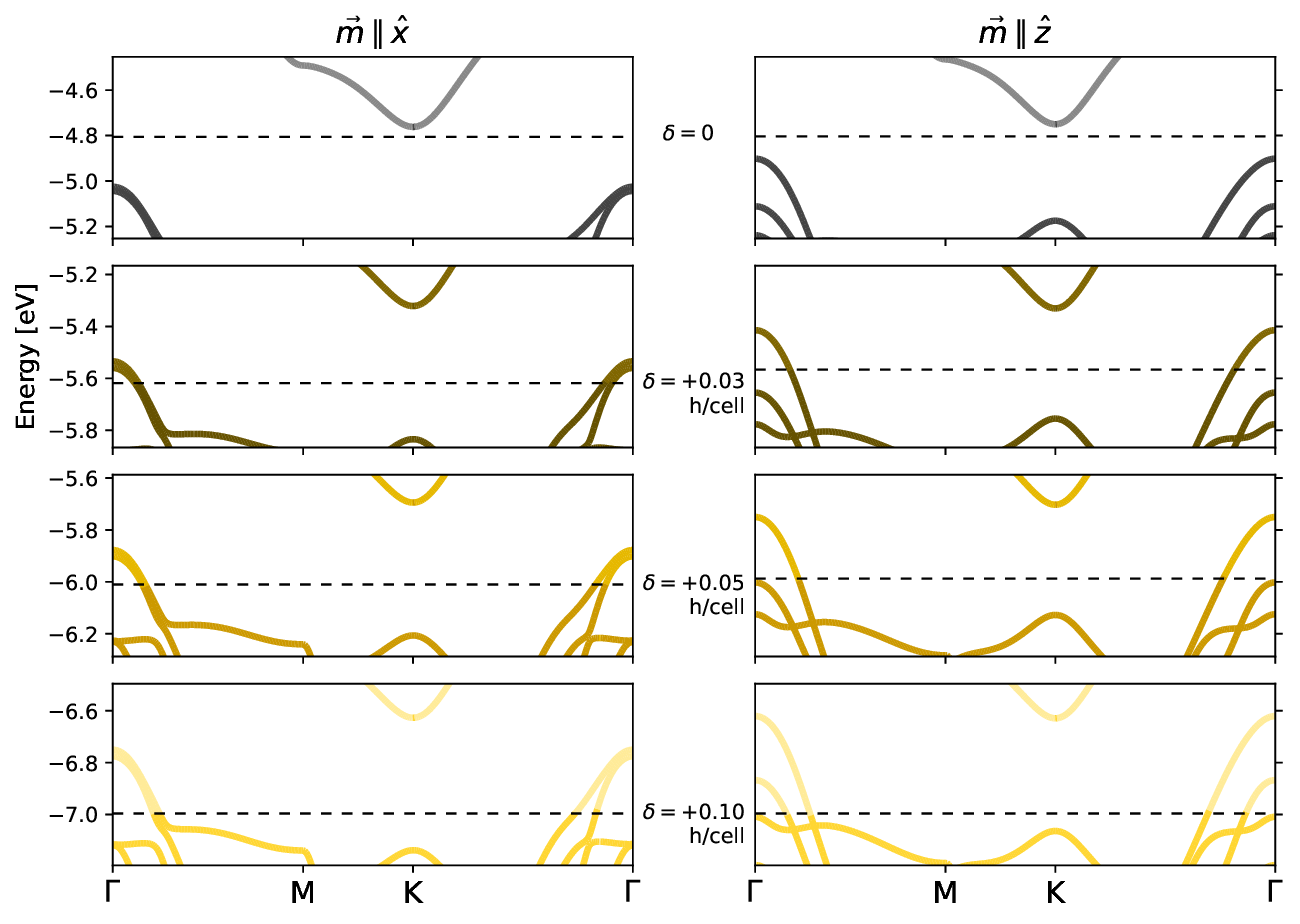} 

    \caption{Fully relativistic band edges of the undoped (top) and the hole-doped (with increasing concentration from $\delta=+0.03$ to $\delta=+0.05$ to $\delta=+0.10$ h/cell towards the bottom panels) \hvte{} for in-plane (left) and out-of-plane (right) magnetization direction. Energy values are relative to the vacuum potential. The horizontal dashed lines are the Fermi levels.}
    \label{fig:bands_soc_doping_u1.3}
\end{figure}

As hole doping lowers $E_F$ to the minority-spin valence bands, the larger energy splitting of the degenerate $d_{xz}$/$d_{yz}$ ($m_l=\pm 1$) states at $\Gamma$ in the OOP magnetization due to SOC results in the removal of higher energy states in the OOP relative to the IP orientation upon hole doping, favoring the stability in the out-of-plane magnetization direction.
The rise in MAE as $\delta$ increases can be attributed to the increasing number of higher energy states emptied in the OOP case compared to the IP case.
Fig.~\ref{fig:bands_soc_doping_u1.3} depicts how the IP and OOP bands change with increasing doping from $\delta=0$ to $\delta=+0.10$ h/cell. A larger splitting is observed around the $\Gamma$ point for the OOP compared to the IP magnetization direction. 
This pushes the topmost valence band to a higher energy level for $\vec{m}\parallel\hat{z}$ compared to $\vec{m}\parallel\hat{x}$. As a result, the energy of the valence states emptied in the OOP case is larger than that in the IP case. This begins to stabilize the OOP direction---increasing the MAE by 1.48~meV from $\delta=0$ to $\delta=+0.03$~h/cell---although the system retains the easy IP magnetization axis overall ($-0.09$~meV MAE). As $E_F$ goes deeper into the valence band when $\delta=+0.05$ and $\delta=+0.10$~h/cell, more higher-energy valence states are emptied in the OOP case which further increases the MAE, switching to a perpendicular MAE of +0.47~meV.

The switching to perpendicular MAE can therefore be attributed to the dependence of the spin-orbit splitting of the degenerate orbitals at VBM on the magnetization direction.
To demonstrate this and to extract general insights, we employ the orbital angular momentum eigenstates $|l,m_l\rangle $.
In the OOP case, the dominant term of  $\hat{L}\hat{S}$,  
$\hat{L}_z\hat{S}_z$, acts strongly on the doubly degenerate minority-spin $| 2, \pm 1\rangle$ state and splits it through $\langle 2, 1 | \hat{L}_z |2, 1 \rangle = \hbar$ in a lower energy level and $\langle 2, -1 | \hat{L}_z |2, -1 \rangle = -\hbar$ in a higher level in the fully relativistic bands. The off-diagonal terms $L_{+}S_{-} + L_{-}S_{+}$ further mix states, increasing the splitting.  
In the IP case, the $\hat{L}_z\hat{S}_z$ term does not dominate anymore and the $\hat{L}_x\hat{S}_x$ or $\hat{L}_y\hat{S}_y$ terms couple weakly the $|2,\pm 1\rangle$ state with other orbitals. The spin-down states $|2,\pm 1 \rangle$ at the VBM are not eigenstates of $\hat{L}_x$, i.e., $ \hat{L}_x |2,\pm 1 \rangle = \frac{\hbar}{2} (\sqrt{6} |2,0\rangle + 2 |2,\pm 2 \rangle) $, so $\langle 2, 1 | \hat{L}_x |2, 1 \rangle = \langle 2, -1 | \hat{L}_x |2, -1 \rangle = 0 $ which means that the degeneracy cannot be lifted by SOC, assuming negligible second-order perturbative coupling to the other states.
Thus, the splitting of the $|2,\pm 1 \rangle$ states is significantly larger when the magnetization axis is OOP.
In general, $\langle l, m_l | \hat{L}_z |l, m_l \rangle = m_l \hbar$ for the OOP magnetization, yielding a first-order spin-orbit splitting for both types of degeneracies ($d_{xz}/d_{yz}$ and $d_{x^2-y^2}/d_{xy}$) and no splitting for $m_l=0$, while the IP magnetization results in zero first-order splittings  $\langle l, m_l | \hat{L}_{x(y)} |l, m_l \rangle = 0 \  \forall \  m_l$.
 This larger splitting in the OOP-magnetization case was also demonstrated in calculations of other systems \cite{smiri2021dft+, daalderop1994magnetic,  sakamoto2021anisotropic}. \added{In Supplementary Section S4, we show that the change in MAE with hole doping can be modeled by a first-order contribution from the energy difference between the OOP and IP valence-band maxima, and a second-order contribution from the energy difference of the emptied bands. Within the rigid-band approximation, this approach yields very good quantitative agreement with the DFT results at small doping levels. These findings strongly support that, in \vte{} and related materials, the MAE is predominantly governed by first-order SOC effects, which directly generate the energy difference between magnetization orientations.}

Previously, the lifting of orbital degeneracies has been related to MCA \cite{daalderop1994magnetic,LesMooHub97,GimCal12}.
In particular, it has been proposed that a $d$ degeneracy can make a large effect on MCA if it is lifted by SOC for only one direction of magnetization (and remains for the other one) and if the degeneracy is located near the Fermi level \cite{LesMooHub97,daalderop1994magnetic}.
The mechanism for the enhancement of MCA (MAE) upon hole doping described in this subsection is related to that, as the hole doping naturally places the Fermi level between the two SOC-split (previously degenerate) $d$-states at the VBM. Regarding the hole-doping-induced changes, a similar mechanism of stabilization of a state/phase of the system with higher-energy valence states upon hole doping has been previously reported for MAE switching in layered (Ba,K)(Zn,Mn)\textsubscript{2}As\textsubscript{2}  \cite{sakamoto2021anisotropic} and for switching from AFM to FM configuration in CrI$_3$ bilayer \cite{GhoStoBin21}.

\subsection{Trends in  MAE and their physical basis in hole-doped \hvx{}}

\begin{figure}[h!]
    \centering
    \includegraphics[width=1\textwidth]{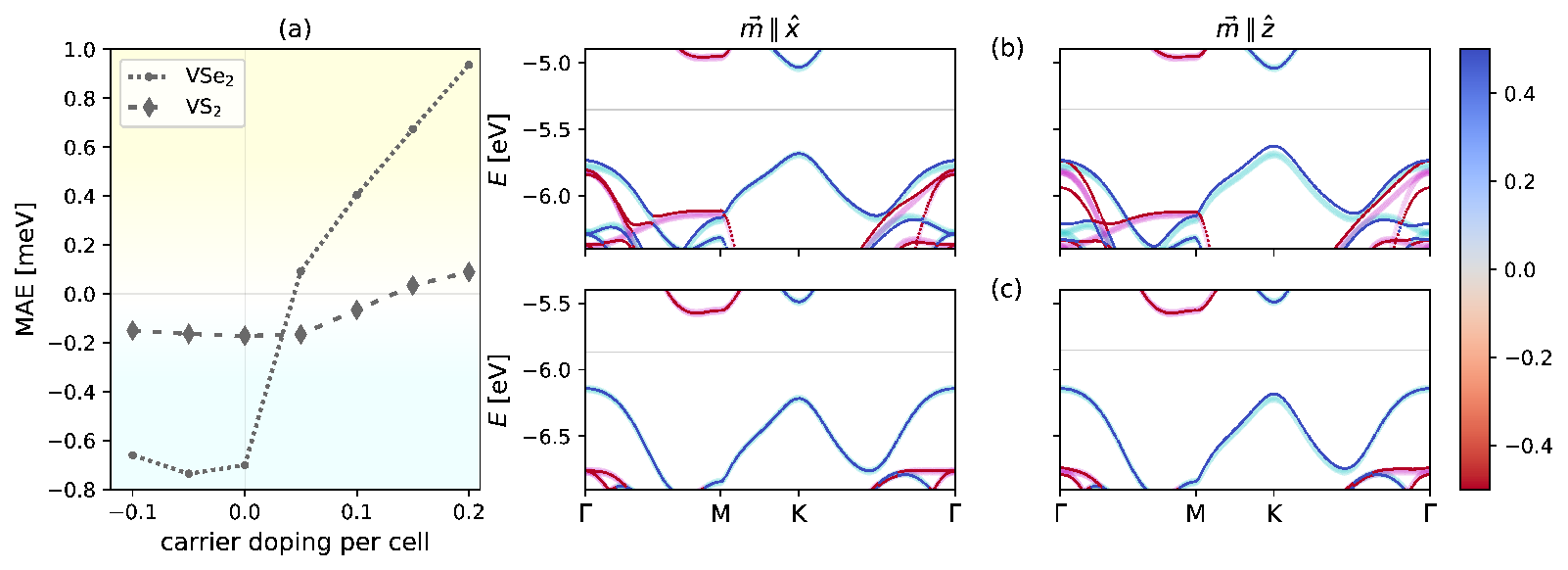}
    \caption{(a) MAE versus charge carrier doping concentrations $\delta$ for FM \vse{} and \vs{}. The calculated $E_{MCA}$ corresponds to the MAE since $E_{SA}$ is negligible (see Supplementary Section~S2).
    Hole doping corresponds to $\delta>0$ while electron doping $\delta<0$.
    \added{Band structure of pristine  (b) \vse{} and (c) \vs{} monolayers without (cyan/magenta thick curves for the majority/minority spin bands) and with (thin curves) spin-orbit coupling effect. The bar at the right indicates the color code for the band projection of the spins along the corresponding magnetization direction. Energy values are relative to the corresponding vacuum potential. The horizontal line marks the Fermi level.}}
    \label{fig:mca_vx2}
\end{figure}

Hole doping of \vse{} and \vs{}  also induces switching to out-of-plane magnetization as shown in Fig.~\ref{fig:mca_vx2}. The pristine \vx{} systems all have easy in-plane axis of magnetization; we calculate the MAE of $-0.17$~meV and $-0.70$~meV for pristine \vs{} and \vse{}, respectively, which agree well with the previously reported DFT values \cite{jafari2023electronic,jiang2024electronic,chen2024first}. Electron doping maintains this preferred magnetization direction while hole doping tends to increase the MAE until sufficient hole doping per cell is reached to achieve PMA.
We estimate by interpolation that the switching to PMA occurs earlier in \vse{} at $\delta=+0.044$~h/cell than in \vs{} at $\delta=+0.135$~h/cell. In both cases, the MAE keeps increasing with further hole doping after switching, reaching the highest value at the largest doping ($\delta=+0.20$~h/cell) that we considered: 0.94~meV for \vse{} and 0.09~meV for \vs{}.

We have shown previously how the switching to PMA in \vte{} can be attributed to the larger SOC splitting of the doubly degenerate $d_{xz}$/$d_{yz}$ VBM at $\Gamma$ in the OOP case.
We turn to the scalar and fully relativistic band structures in Fig.~\ref{fig:mca_vx2} to investigate if similar magnetization-axis-dependent SOC effect occurs at the VBM---and/or nearby valence states---in \vse{} and \vs{}.
The band structures of the three vanadium dichalcogenides have some common features.
All three are semiconductors, with the energy gap of \vse{} ($E_g=0.65$~eV) and \vs{} ($E_g=0.57$~eV) being larger than that of \vte{}.
For the mechanism of hole-doping-induced PMA, we focus on the highest valence band. \added{In  \vse{} and \vs{}, there are two valence band maxima close in energy, at $\Gamma$ and  $K$.} 
The highest state at $\Gamma$ changes orbital character from $d_{xz}$/$d_{yz}$ in \vte{} to $d_{z^2}$ in both \vse{} and \vs{}, as can be seen in the orbital-projected bands for \vse{} and \vs{} in Supplementary Section~S5.
The highest state at the $K$ point in all three cases is singly degenerate having $d_{xy}$/$d_{x^2-y^2}$ character.
While in \vte{} and \vs{} the VBM is at $\Gamma$, in \vse{} it is shifted to K. The $d_{z^2}$ orbital does not exhibit first-order spin-orbit splitting, as can be seen from the comparison of scalar-relativistic and fully relativistic bands in Fig.~\ref{fig:mca_vx2}.  Thus, in the fully relativistic case its energy does not depend on the magnetization orientation, yielding zero first-order contribution to MAE.

The state at K with in-plane orbital character is the VBM in \vse{} and a local VBM in \vs{}. The $d_{xy}$ and $d_{x^2-y^2}$ can form $|2,\pm 2\rangle$ eigenstates of $\hat{L}_z$ with eigenvalues $m_l=\pm2$ contributing to the $\hat{L}_z\hat{S}_z$ spin-orbit effect which raises the (local) VBM higher when the spins are aligned along $z$ than along $x$. 
It exhibits a significant SOC-induced energy shift given by $\Delta \mathcal{E}^{\perp/\parallel}=\mathcal{E}^{\perp/\parallel}_{\textrm{with SOC}} - \mathcal{E}^{\perp/\parallel}_{\textrm{no SOC}}$, each calculated relative to the vacuum potential. 
For \vse{} the shift $\Delta \mathcal{E}^{\perp}_K=64$~meV is sizable and much larger than $\Delta \mathcal{E}^{\parallel}_K=7$~meV. Thus, under even a small hole doping, the larger decrease in energy is immediately obtained for OOP magnetization direction, relative to IP. Indeed, the MAE in Fig.~\ref{fig:mca_vx2} switches to PMA already at the smallest hole doping we considered ($\delta=+0.05$~h/cell) and keeps increasing as more states are emptied from the peak at K.
We note that the reported valley splitting \cite{wang2021effects,jiang2024electronic,chen2024first} at K due to SOC when $\vec{m}\parallel \hat{z}$ does not affect the mechanism for PMA switching in \hvx{}: see Supplementary Section~S6.

\begin{figure}[hbt]
    \centering    
    \includegraphics[width=1\linewidth]{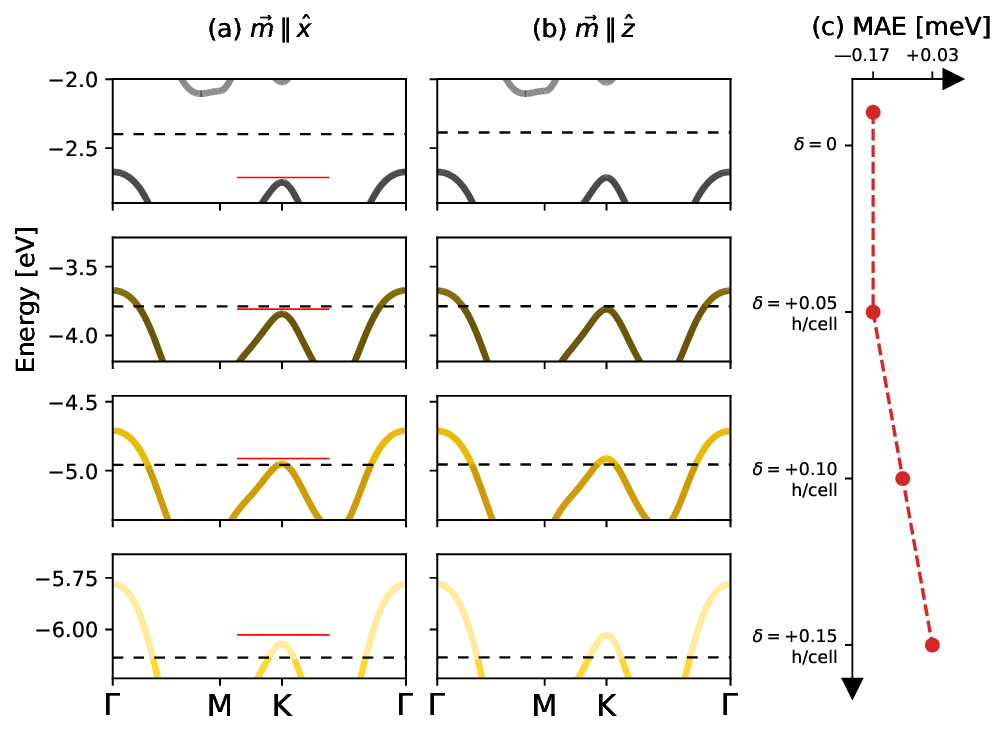} 
    \caption{Fully relativistic band edges of the undoped (top) and the hole-doped FM \hvs{} (with increasing concentration from $\delta=+0.05$ to $\delta=+0.10$ to $\delta=+0.15$ h/cell towards the bottom panels) for (a) in-plane and (b) out-of-plane magnetization direction. Energy values are relative to the vacuum potential. The horizontal dashed lines are the Fermi levels. \added{The red horizontal bar marks the local VBM at K when $\vec{m}\parallel\hat{z}$}  (c) Doping level vs. corresponding MAE.} 
    \label{fig:bands_soc_doping_u1.3_VS2}
\end{figure}

In the case of \vs{}, the initial hole doping does not increase MAE, as the emptied VBM states ($|2,0\rangle$) at $\Gamma$ are not affected by spin-orbit and remain at the same energy for the OOP and IP magnetization directions.
This can be seen in the first two panels of Fig.~\ref{fig:bands_soc_doping_u1.3_VS2} which displays the bands of hole-doped \vs{} close to $E_F$ for two magnetization directions at different hole-doping levels.
It is evident that increasing the doping concentration from $\delta=0$ to $\delta=+0.05$~h/cell does not significantly change the MAE ($-0.17$~meV in both cases), as shown in  Fig.~\ref{fig:bands_soc_doping_u1.3_VS2}.
Only when the $|2,\pm 2\rangle$ states at K are depleted starting at $\delta=+0.10$~h/cell does the MAE begin to increase until it switches to PMA at $\delta=+0.135$~h/cell. As in \vse{}, we observe a greater shift in energy due to SOC in the highest valence state at K when spins are OOP ($\Delta \mathcal{E}_K^\perp =  36$~meV) than IP ($\Delta \mathcal{E}_K^\parallel =  1$~meV).
Therefore, as shown in Fig.~\ref{fig:bands_soc_doping_u1.3_VS2}c, the MAE responds differently to two classes of valence band maxima—those with vanishing first-order spin-orbit coupling ($d_{z^2}$ orbital) and those with pronounced first-order spin-orbit splitting/shift ($d_{xy}/d_{x^2-y^2}$ in \vs{} and $d_{xz}/d_{yz}$ in \vte{})—demonstrating the critical role of orbital degeneracies and spin-orbit interactions in VBM  for hole-doping-induced PMA.

Thus far, we have used the orbital properties at  VBM and at nearby energies to explain the transition from in-plane MAE to PMA upon hole doping. Next, we investigate the fundamental mechanisms governing the orbital character and energy alignments, as well as their evolution from \vte{} to \vs{}.
Crystal field effects, orbital hybridization, and exchange splitting influence  MAE under hole doping primarily by modifying the ordering of valence states, including the VBM. In contrast, spin-orbit coupling has a more direct impact: it not only affects the band ordering but it also determines the splitting of degenerate states which results in stabilization of  PMA,  controlling the critical hole doping concentration for the transition.

\begin{figure}[h!]
    \centering
    \includegraphics[width=\linewidth]{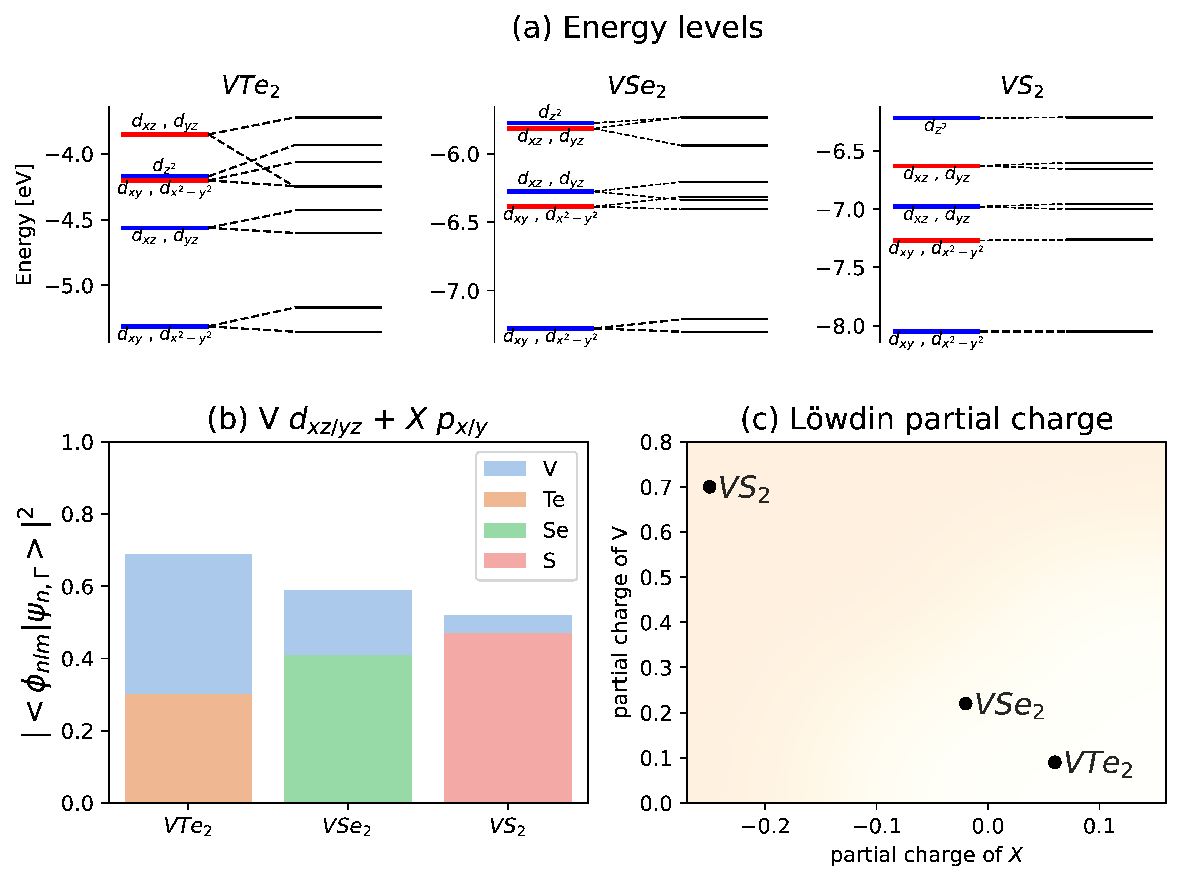}
    \caption{ (a) Vanadium energy levels at $\Gamma$ without (blue: majority spin, red: minority spin) and with (black) spin-orbit coupling effects in the case of $z$-oriented magnetization. Energy values are relative to the corresponding vacuum potential.
    \added{Trends in hybridization of the minority-spin vanadium $d_{xz}/d_{yz}$ and chalcogen $p_x/p_y$ orbitals at $\Gamma$ (b) and in vanadium/dichalcogenide L\"owdin partial  charges (c) in \vx{} ($X=$ Te, Se, and S). In (b), the weight of the orbital character is determined by projection coefficients $|\langle \phi_{n l m} | \psi_{n, \Gamma}\rangle|^2$ onto atomic states at $\Gamma$ in \vx{}. A larger projection value on V (indicated by the height of the blue bar for each compound) and smaller on $X$ indicate lower hybridization.  In  (c), a more positive partial charge on V and more negative on $X$ imply greater ionicity on the V-$X$ bond.}}
    \label{fig:splittings_Gamma}
\end{figure}

We first analyze the trends in splitting/shift of VBM.
The trend of decreasing MAE of the hole-doped \vx{} for $X$ changing from Te to Se to S---observed also in previous studies \cite{fuh2016newtype, jafari2023electronic}---follows the trend of decreasing atomic SOC from Te to S.
This can be expected since SOC is responsible for the splitting/shift of the VBM in the OOP case which directly influences the magnitude of MAE.
To compare the SO effects of the valence states in the vanadium dichalcogenides, we schematically plot in Fig.~\ref{fig:splittings_Gamma}a both scalar-relativistic and fully relativistic energy states at  $\Gamma$,  which transform under the full symmetry group of the monolayers. 
Twofold degeneracies can be found by symmetry at $\Gamma$ in states with $d_{xz}/d_{yz}$ ($|2,\pm  1\rangle$)  and $d_{xy}/d_{x^2-y^2}$ ($|2,\pm 2\rangle$) character.
The fully relativistic levels clearly show that the effect of SO interaction is more pronounced in \vte{} due to the heavier Te atom; in contrast, the SO splittings in \vs{} are much smaller.  
For the large SO splitting of \vte{} it is crucial that the VBM has a rather strong Te character.
This is shown in Fig.~\ref{fig:splittings_Gamma}b in which we compare the weights of the vanadium/dichalcogenide character of the V $d_{xz}/d_{yz}$+ $X$ $p_x/p_y$  state at $\Gamma$ in \vx{}, calculated as the state's projection onto the given orbitals. The dichalcogenide character of this state is increasing from Te to Se to S.
However, in \vs{}—where the dichalcogenide character of the VBM at $\Gamma$ is the most  pronounced—the spin-orbit splitting remains relatively small. This arises because the atomic SOC of S $p$-orbitals is only somewhat stronger than that of V $d$-orbitals \cite{NIST_ASD}.
Thus, the $d_{xz}$/$d_{yz}$ minority-spin states are split due to SOC by 337~meV in \vte{}, 206~meV in \vse{}, and 54~meV in \vs{}: see Fig.~\ref{fig:splittings_Gamma}a.
This trend also applies to the SOC-induced energy shift in the highest valence state at K with $\Delta \mathcal{E}_K^\perp =  124$~meV in \vte{}, $\Delta \mathcal{E}_K^\perp =  64$~meV in \vse{}, and $\Delta \mathcal{E}_K^\perp =  36$~meV in \vs{}.

In order to understand the trend in the orbital character of VBM at $\Gamma$ and its energy alignment, we examine the crystal field properties and the effect of exchange splitting. 
In \hvx{}, the degeneracy of the V $d$-orbitals is broken by the crystal field due to the trigonal prismatic coordination.
The effect of this crystal field can be described by the energy difference at $\Gamma$ between the $d_{xy}$/$d_{x^2-y^2}$ (E') and $d_{xz}$/$d_{yz}$ (E") minority-spin states, which is the smallest for \vte{} (325~meV) and the largest for \vs{} (636~meV): see Fig.~\ref{fig:splittings_Gamma}a.
This implies an increasing crystal field splitting from \vte{} to \vs{}.
A summary of the energy splittings and shifts can be found in Supplementary Table~S3.
Out-of-plane $d$-orbitals, especially $d_{z^2}$, experience larger energy shifts in contrast with the in-plane $d$-orbitals that feel the least repulsion from the ligands.
Consequently, a stronger crystal field is expected to elevate the $d_{z^2}$ state at $\Gamma$ relative to the other $d$-states. Indeed,  \vs{}---with the largest crystal field splitting---has the $d_{z^2}$ state at the highest position at $\Gamma$, while the other states are significantly lower in energy. In this case, the strong crystal field is a result of stronger ionic character of the vanadium-dichalcogenide bond. 
Fig.~\ref{fig:splittings_Gamma}c shows that the V-S bond has the greatest ionicity among the \vx{} systems, measured by the L\" owdin partial charge on V and X. It can be seen that V has the largest partial charge ($+0.70$) in \vs{}, and S has the most negative partial charge ($-0.25$). This is consistent with the trend of increasing electronegativity from Te to Se to S.

The exchange splitting has a similar effect of favoring the majority $d_{z^2}$ as VBM. The exchange splitting is the smallest for \vs{} and the largest for \vte{}. In \vte{}, it is larger than the crystal field splitting between majority-spin A' and E'' states, pushing the minority-spin $d_{xz}$/$d_{yz}$ to the VBM.
In \vse{} and \vs{}, the exchange splitting between E'' states is smaller than the A'-E'' crystal field splitting, which places the majority-spin $d_{z^2}$ orbital at a higher energy level than minority-spin $d_{xz}$/$d_{yz}$ at $\Gamma$.
This is evident in Fig.~\ref{fig:splittings_Gamma}a where the scalar-relativistic energy levels depict trends in exchange and crystal field splitting at $\Gamma$. The energy separation between the majority and minority spin $d_{xz}$/$d_{yz}$ states at $\Gamma$ goes from 655~meV in \vte{} to 350~meV in \vse{} to 35~meV in \vs{}.
This follows from the trends in chalcogen atomic radius---and the equilibrium lattice parameter decreasing from \vte{} to \vs{}---and orbital hybridization.
The relative dichalcogenide character shown in Fig.~\ref{fig:splittings_Gamma}b serves as a direct metric for  V $d$–X $p$ orbital hybridization, obtained through atomic-orbital-projected band structure calculations.
It shows increasing hybridization from \vte{} to \vs{} at $\Gamma$. This is confirmed by comparing overall energy overlap and contribution weights of relevant orbitals, as well as the bandwidths in the projected density of states (PDOS) across \vx{}, included in Supplementary Section~S7.

\added{The increasing hybridization and decreasing exchange splitting from \vte{} to \vs{} imply a trend of decreasing magnetic moment $m$ of vanadium in \vx{}. For instance, the vanadium magnetic moments in the pristine structures are as follows: $m_{VTe_2}=1.25\mu_B>m_{VSe_2}=1.07\mu_B > m_{VS_2}= 0.96\mu_B$. A more detailed discussion on the trends of magnetic moment upon hole doping can be found in Supplementary Section~S7.3. We also show in Supplementary Section S8 that, while the charge-doping-induced changes in magnetic moment are sensitive to the Hubbard-$U$, the switching to PMA upon hole doping occurs regardless of the choice of $U$: ${\rm MAE}>0$ at $\delta=+0.05$~h/cell for $U=$ 0, 1.3, and 1.7 eV. This can be attributed to the doubly degenerate VBM states at $\Gamma$ that persist even with increasing $U$.}

\subsection{General conditions for MCA enhancement upon hole doping and material examples}

So far, we showed that essentially the same mechanism of PMA upon hole doping applies to three main representatives of H-VX$_2$, with differences in orbital character of the SOC-split/shifted orbitals.
The mechanism is based on a larger decrease in energy upon hole doping for the OOP magnetization direction due to a greater (first-order) spin-orbit splitting/shift of the $d_{xz}/d_{yz}$ or $d_{x^2-y^2}/d_{xy}$  mixed orbitals at VBM or at nearby energies.
Clearly, for this mechanism to function, the VBM (or degenerate states close in energy) must possess atomic character with appreciable SOC.
So, we will focus on magnetic semiconductors with such VBM.

For the structures with crystallographic point groups that preserve the $d_{xz}$/$d_{yz}$ and $d_{x^2-y^2}$/$d_{xy}$ degeneracies, the splitting/shift is larger in the OOP than in the IP direction for all $m_l\ne 0$ $d$-orbitals---both of the degenerate sets yield larger splitting in the OOP case. For the $d_{z^2}$ orbital, the first-order terms are vanishing and the second-order terms determine whether OOP or IP direction is favored upon hole doping, but the contributions are small relative to the first-order splittings/shifts of the other orbitals.
Indeed, for all three \hvx{} compounds studied in this work, the second-order terms are negligible relative to the splitting/shift of the bands with $m_l\ne 0$ character, as can be seen from the splittings shown in Fig.~\ref{fig:splittings_Gamma}.
The simultaneous $d_{xz}$/$d_{yz}$ and $d_{x^2-y^2}$/$d_{xy}$  degeneracies are found in a number of point groups, which we list in Table~S4 in SI. The given point groups describe crystallographic symmetry before considering relativistic effects.

The H-phase monolayers belong to the $D_{3h}$ point group, where  $d_{xz}$/$d_{yz}$ and $d_{x^2-y^2}$/$d_{xy}$ degeneracies  exist only at the $\Gamma$ point. However, it can be observed that throughout the Brillouin zone, the $d_{xz}$ and $d_{yz}$ (or $d_{x^2-y^2}$ and $d_{xy}$) are mixed, which is a consequence of the $C_3$ symmetry which mixes the orbitals from degenerate pairs.
Only at $\Gamma$ and K are the weights of the two degenerate states identical, but at the other k-points, the states have varying weights of the two orbital characters in a singly-degenerate state.
As we saw in the case of \vse{}  with the VBM at the K point, also singly degenerate states can be (strongly) affected by SOC and can contribute to the enhancement of MAE by shifting to higher energy in the case of OOP magnetization. 
Thus, as long as VBM (or states close in energy) consists of $d$-states with $m_l\ne0$, we can expect that the hole doping will initiate a positive trend in MAE.
However, a broader range of symmetry classes---including tetragonal, hexagonal, and cubic systems---can selectively maintain the degeneracy of the $d_{xz}$/$d_{yz}$ orbitals while splitting other $d$-orbital states; these are shown in Table~S5. Also \added{the $d_{xz}$/$d_{yz}$ orbital degeneracy is expected} to enable targeted band engineering to position these degenerate orbitals at or near the VBM, promoting a positive MAE change upon hole doping, \added{as will  be shown in the next subsection}.
Likewise in the case of $p$-orbitals at the VBM, the degenerate $p_x/p_y$ orbitals are expected to yield a larger spin-orbit splitting/shift in the OOP case (occurring in cubic, hexagonal, and tetragonal symmetry). The point groups preserving $p_x/p_y$ degeneracy are listed in Tables~S4 and S5.

Our analysis implicitly assumes a system with an energy gap.
While in principle a similar analysis might be applied in the case of a metal, the practical implementation could prove significantly more complex due to the multiple states at the Fermi energy and strong many-body interactions.  

The primary limitation of our analysis arises in cases of strong band renormalization upon doping. 
\replaced{Indeed, as the system starts deviating from the rigid band model, the quantitative prediction of MCA trends under hole doping becomes less precise. Strong band renormalization can cause the correlation between the MCA change and the total energy of the emptied valence states to deviate from linearity.
Nevertheless, the rigid band approximation is not strictly necessary to capture the overall increasing trend of MCA upon hole doping, which can still be understood in terms of the enhanced spin–orbit splitting near the valence band edge.}
{Indeed, as the system deviates from the rigid band model, the ability to predict MCA trends under hole doping becomes less reliable. Nevertheless, the rigid band approximation is not strictly necessary to capture the positive correlation between hole doping and enhanced MCA, particularly in the low-doping regime.}
Some mild band renormalization is not going to influence qualitatively the trend, but it can shift the onset of PMA.

Overall, the conditions for the enhancement of MCA under hole doping in a magnetic semiconductor are rather intuitive and simple: (i) VBM consisting of degenerate $d$ and/or $p$ orbitals or mixed orbital character states with $m_l\ne 0$  (the existence of degenerate orbitals---not necessarily at the VBM---is ensured for crystallographic point groups given in Tables~S4 and S5) (ii) the VBM states should possess atomic character with finite atomic SOC. 
The first condition based on the degeneracy of the $d$ and $p$ orbitals allows a large number of point groups, but it should be noted that this condition requires a particular position (at the VBM or at nearby energies) of the degenerate states and therefore a correct point group does not guarantee the enhancement of MCA under hole doping. Moreover, majority of the systems satisfying these conditions will be magnetic 2D systems.  Although bulk systems are symmetry-allowed by our conditions, few intrinsic magnetic semiconductors exist in bulk form. Apart from rare-earth-based materials, research primarily focuses on diluted magnetic semiconductors and van der Waals-coupled 3D systems, many of which crystallize in the point groups listed in Tables~S4 and S5.

\replaced{Here are some examples}{\textbf{Examples}} of materials that satisfy the above conditions and exhibit—or are theoretically predicted to exhibit—an increase in MCA upon hole doping:

\textit{I. $D_{3h}$ point group:} Ab initio studies demonstrated the switching from IP to OOP magnetization upon hole doping in other FM semiconductors such as  ScI\textsubscript{2} \cite{wu2023realizing} \textit{(Example 1)} and  Os-doped MoTe\textsubscript{2} \cite{torun2015tuning} \textit{(Ex.~2)}.
Both examples satisfy the above conditions for enhancement of MCA: \textit{Ex.~1} has degenerate Sc-$d_{xy}$/$d_{x^2-y^2}$ orbitals at the VBM while \textit{Ex.~2} has Os-$d_{xz}/d_{yz}$ and $d_{xy}$/$d_{x^2-y^2}$, and both have atomic character with strong SOC (from the I- and Te-$p_x$/$p_y$ orbitals) at the VBM.
In monolayer FeCl\textsubscript{2} \cite{saritas2022piezoelectric} \textit{(Ex.~3)} hole doping further increases the already positive MCA, owing to the degenerate Fe $d_{xy}$/$d_{x^2-y^2}$ orbitals at the VBM.
We expect that the MCA in other materials such as  VSi\textsubscript{2}N\textsubscript{4} \cite{cui2021spin} \textit{(Ex.~4)} and  FeI\textsubscript{2}  \cite{saritas2022piezoelectric} \textit{(Ex.~5)} could also be increased upon hole doping as they possess degenerate states (V-$d_{xy}$/$d_{x^2-y^2}$ in the former and I-$p_x$/$p_y$ in the latter) which also provide finite SOC at the VBM---this could be checked by DFT calculations or measured experimentally.

\textit{II. $C_{3v}$:} In Janus H-V$XY$ ($X\neq Y=$ Te, Se or S) monolayers \textit{(Ex.~6)}, the $d_{xz}/d_{yz}$ and $d_{xy}$/$d_{x^2-y^2}$ VBM degeneracies can be found in VSeTe at $\Gamma$ and in VSSe at K, respectively, which enable PMA upon hole doping as shown in recent first-principles calculations \cite{guan2020predicted, li2022two}.

\textit{III. $D_{3d}$:} The  two-dimensional CrI\textsubscript{3} \textit{(Ex.~7)} already has perpendicular magnetization, but Kim et al \cite{kim2019exploitable} showed that a small hole doping can increase its MCA. This material has degenerate I $p_x$/$p_y$-states at the VBM at $\Gamma$, satisfying the listed conditions. 

\textit{IV. $D_{2d}$:} Experimentally, PMA was observed in a layered (Ba,K)(Zn,Mn)\textsubscript{2}As\textsubscript{2} \textit{(Ex.~8)} ferromagnetic semiconductor which was attributed to the doped holes residing in one of the SOC-split $d_{xz}/d_{yz}$ states \cite{sakamoto2021anisotropic}. With the many materials databases available, new materials can be screened and identified for possible enhancement of MCA upon hole doping. For example, we found in Computational 2D Materials Database \cite{haastrup2018computational} a class of dynamically stable FM materials \textit{(Ex.~9)}: namely Mn$X_2$ ($X=$ I, Br, Cl). They have two-fold Mn-$d_{xz}/d_{yz}$ (hybridized with $X$-$p_x$/$p_y$) degeneracy at the VBM at $\Gamma$ which makes them candidates for increasing MAE upon hole doping. We performed DFT calculations and found that, for example in MnBr\textsubscript{2}, indeed $E_{MCA}$ increases from 0.2 to 1.5~meV upon $\delta=+0.05$~hole doping per cell. The calculations and discussion of the mechanism for the hole-doping-induced MCA change in these types of materials are presented in Supplementary Section~S10.

\subsection{Band engineering for perpendicular MAE upon hole doping}

We have established the importance of the degenerate $|l=2,m_l\neq0 \rangle$ states at the VBM in switching to PMA at minimal doping levels. This is exemplified by the significant increase in MAE from $\delta=0$ to $\delta=+0.05$ h/cell in \vte{} (VBM: $|2,\pm1\rangle$ at $\Gamma$) and \vse{} (VBM: $|2,\pm2\rangle$ at K), but not in \vs{} (VBM: $|2,0\rangle$ at $\Gamma$). In \vs{} we observe an MAE increase from $\delta=+0.05$ to $\delta=+0.10$ h/cell only when the hole doping concentration is sufficiently high to empty the $d_{xy}/d_{x^2-y^2}$ ($m_l=\pm2$) valence states at K: see Fig.~\ref{fig:bands_soc_doping_u1.3_VS2}. This begs the question if the system can be modified to tailor the bands placing the degenerate $d$-orbitals with $m_l\neq0$ at the VBM, for example in \vs{}. We emphasize that the identification of these degenerate states at the valence band edge can be done at the scalar-relativistic level---which does not require high computational cost---in predicting doping-induced changes in MAE after band engineering. 
It could be expected that the desired effect of lowering the energy of the $d_{z^2}$ state at $\Gamma$ to below the energy of the  $d_{xy}/d_{x^2-y^2}$ orbital at K in the valence band edge could be obtained by compressive in-plane strain which destabilizes the in-plane orbitals and stabilizes the $d_{z^2}$ state.

\begin{figure}[h!]
    \centering
    \includegraphics[width=1\linewidth]{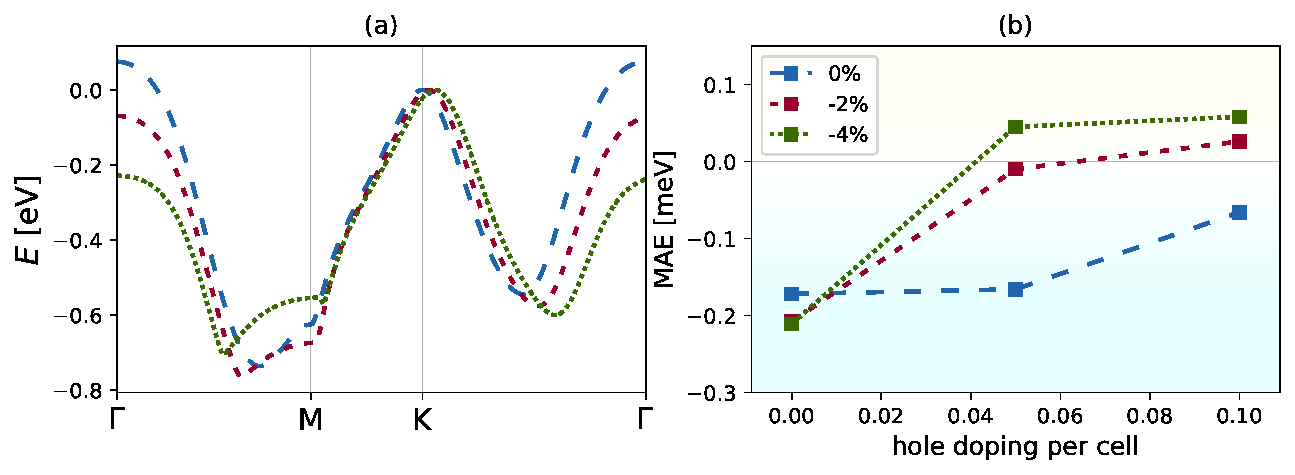}
    \caption{(a) Valence band edge and (b) MAE versus hole doping concentrations $\delta$ with increasing biaxial compressive strain in \vs{}. The highest valence state at K is set to zero in the band edge for each corresponding strain. The dashed blue line corresponds to the pristine case, the dashed red to the strain of --2\%, and dotted green to the strain of --4\%. The V states  at $\Gamma$ have  the  $d_{z^2}$ character and at K, $d_{x^2-y^2}/d_{xy}$.}
    \label{fig:vs2-strain-dope}
\end{figure}

Thus, to demonstrate how the valence bands can be engineered to promote PMA by minimal hole doping, we introduce biaxial compressive strain in \vs{} to shift the VBM from  $|2,0 \rangle$ at $\Gamma$ to $|2,\pm2 \rangle$ at K. Fig.~\ref{fig:vs2-strain-dope}a shows that  biaxial compressive strain indeed pushes the $d_{z^2}$ state at $\Gamma$ to a lower energy level than the $d_{xy}/d_{x^2-y^2}$ orbital at K in the valence band edge. In the unstrained \vs{}, the VBM has $d_{z^2}$ character at $\Gamma$ which is unaffected by SOC when rotating the magnetization axis while the VBM in the strained system has $d_{xy}/d_{x^2-y^2}$ character at K, which can be shifted by SOC. This yields a larger increase in MAE when doping the strained \vs{} even with small hole concentration.

Without strain, there is no significant change in the MAE from $\delta=0$ to $\delta=+0.05$ h/cell, and the MAE becomes positive only when $\delta=+0.135$~h/cell: see Fig.~\ref{fig:bands_soc_doping_u1.3_VS2}.
But Fig.~\ref{fig:vs2-strain-dope}b demonstrates that larger compressive strain not only amplifies the enhancement of MAE,  but also reduces the critical hole doping concentration required to induce the transition to PMA.
With $-2\%$ compressive strain, the MAE at $\delta=+0.10$~h/cell is 0.026~meV while with $-4\%$ compressive strain, the MAE becomes positive already at $\delta=+0.05$~h/cell with $+0.045$~meV MAE.

Other chemical or structural modifications can then be explored in \hvx{} and related systems that will push degenerate $|l=2,m_l \neq0\rangle$ states to the VBM, promoting perpendicular MCA even for small hole doping.

\section{Conclusions}

H-phase vanadium dichalcogenide monolayers \vx{} ($X$=Te, Se, and S) are ferromagnetic semiconductors with easy in-plane magnetization that can be switched to the out-of-plane magnetization axis by hole doping. With our study centered on \hvx{}, we found that the mechanism behind the switching from in-plane to perpendicular magnetic anisotropy upon hole doping is governed by the larger spin-orbit splittings and shifts when the magnetization axis is in the OOP direction.  Depletion of the resulting higher-energy states in the case of hole doping reduces energy more in the OOP case, and---at some concentration of holes---switches to PMA. Since the first-order spin-orbit splitting arises from orbital degeneracies at the topmost valence states, such mechanism is transferable to other systems where symmetries enforce such degeneracies. \added{Importantly, this mechanism provides meaningful insights also when conventional approaches, such as the force theorem or second-order perturbation theory, fail to provide a reliable description---for instance, in low-dimensional systems with strong spin–orbit splitting.}  

The decreasing trend in the magnitude of the magnetic anisotropy from \vte{} to \vs{} follows the decreasing atomic SOC from Te to Se to S. The vanadium magnetic moment also exhibits a decreasing trend which follows from the increasing hybridization and the decreasing exchange splitting. We demonstrated how the increasing crystal field splitting, decreasing exchange interaction, and increasing orbital hybridization affect the highest valence states in \vte{}, \vse{}, and \vs{}.

Although the spin-orbit splittings and shifts can only be observed at the fully relativistic level, we have established a set of straightforward and simple criteria to enhance MAE under hole doping in other semiconducting magnetic materials. These criteria are based on orbital degeneracies or state mixing ($d$ and/or $p$ orbitals with $m_l\neq0$) possessing atomic contributions with finite SOC at the VBM or at nearby energies. We have identified at least 21 point groups preserving the said degeneracies. Moreover, we presented a few examples of previously reported systems enhancing MCA with hole doping that satisfy our conditions and some systems which we consider to be highly likely to display a trend of increasing MCA with hole doping.

Our mechanism guided by the general criteria mentioned allows targeted engineering of the band structure to increase MAE and shift the onset of PMA to minimal hole doping levels, as we have demonstrated here for \vs{}. Leveraging on crystallographic symmetries and easily accessible electronic properties within the scalar-relativistic framework, our criteria can be systematically implemented in a high-throughput screening of materials targeted for increasing MAE by hole doping.

\begin{acknowledgments}
We are grateful to Nadia Binggeli  for insightful discussions. 
\end{acknowledgments}

\section{Data availability}
The data that supports the ﬁndings of this article is not
publicly available. The data is available from the authors upon reasonable request.

\putbib[references.bib]
\end{bibunit}


\clearpage
\newpage

\begin{bibunit}[apsrev4-2]

\begingroup

\makeatletter
\def\@hangfrom@section#1#2{#2}
\pretocmd{\@bibitem}{\raggedright}{}{}
\setcounter{section}{0}
\renewcommand \thesection{S\@arabic\c@section}
\setcounter{table}{0}
\renewcommand\thetable{S\@arabic\c@table}
\setcounter{figure}{0}
\renewcommand \thefigure{S\@arabic\c@figure}
\makeatother

\newgeometry{margin=0.5in}
\small
\setlength{\parindent}{15pt}
\setlength{\parskip}{0pt}
\singlespacing

\noindent\Huge{\textbf{Electronic Supplementary Information:}}

\noindent\Large{\textbf{Transferable mechanism  of perpendicular magnetic anisotropy switching by hole doping in VX\textsubscript{2} (X=Te, Se, S) monolayers}}

\vspace{0.3cm}
\noindent\large{John Lawrence Euste,\textit{$^{a,b}$} Maha Hsouna,\textit{$^{a,b}$} and Nata\v sa Stoji\' c \textit{$^{b}$}}

\small{\textit{$^a$~Scuola Internazionale Superiore di Studi Avanzati (SISSA), Trieste, Italy.}}

\small{\textit{$^b$~The Abdus Salam International Centre for Theoretical Physics (ICTP), Trieste, Italy.}}

\section{Computational details and system properties}
\label{section:comp_details}
DFT calculations were performed using the parameters listed in Table \ref{tab-a:params} for each H-phase monolayer (\vte{},\vse{}, and \vs{}) investigated in our study. 
For all three systems, a Gaussian broadening of 0.005 Ry was used with a Brillouin zone sampling grid of $26\times26\times1$ to calculate the MCA of the pristine and charge-doped systems.
These parameters were carefully chosen to ensure that the MCA values are accurate within an error of less than 0.03~meV.

\begin{table}[h!]
\caption{Input parameters for the DFT calculations.}
\centering
\begin{tabular}{l l l l}\toprule
Parameter & \vte{} & \vse{} & \vs{} \\\midrule
kinetic energy [Ry] cutoff for wavefunctions & 60 & 80 & 60 \\
kinetic energy [Ry] cutoff for charge density & 650 & 800 & 640 \\
convergence threshold [Ry] for self-consistency & $7\times 10^{-10}$ & $7\times 10^{-10}$ & $1\times 10^{-13}$ \\ \bottomrule
\end{tabular}
\label{tab-a:params}
\end{table}

For phonon calculations, dynamical matrices were calculated on a $4\times4\times1$ $q$-point grid with a $10^{-16}$~Ry convergence threshold. Imaginary frequencies were not found in the phonon dispersion of \hvte{} in Fig. \ref{fig:phonon-vte2} indicating the stability of the H-phase.

\begin{figure}[h!]
    \centering
    \includegraphics[width=0.3\linewidth]{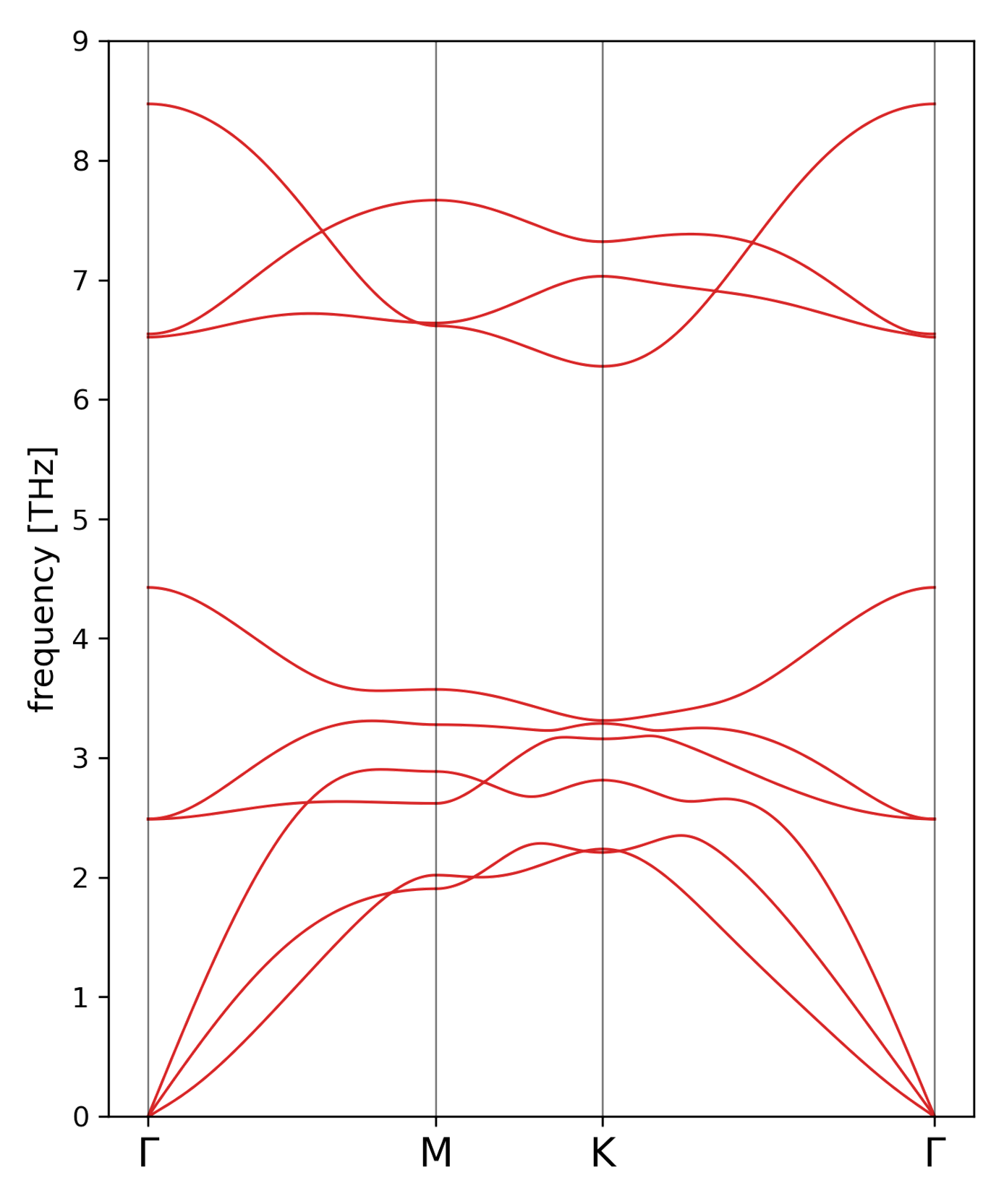}
    \caption{Phonon dispersion of \hvte{}}
    \label{fig:phonon-vte2}
\end{figure}

\newpage
\section{Shape anisotropy}
\label{section:E_SA}

The shape anisotropy energy $E_{SA}$ arises from magnetic dipole-dipole interaction, given by:

\begin{equation}
    E_{SA}=\frac{\mu_0}{8\pi} \sum_{j\ne i} \left[\frac{{\mathbf{m}}_{i} \cdot {\mathbf{m}}_{j}}{{r}_{ij}^3}-\frac{3 ({\mathbf{m}}_{i} \cdot {\mathbf{r}}_{ij})({\mathbf{m}}_{j} \cdot {\mathbf{r}}_{ij})} {{r}_{ij}^5} \right]
    \label{eq:Edip}
\end{equation}

In \vx{}, $\mathbf{m}_i=\mathbf{m}_j$ for V atoms and $\mathbf{r}_{ij}=a$ where $a$ is the corresponding lattice constant. With these, the relation between the $E_{SA}$ of two phases or configurations is simplified to:

\begin{equation}
    {E_{SA_2}} = \left( \frac{a_1}{a_2} \right)^3 \left( \frac{m_2}{m_1} \right)^2 {E_{SA_1}}
    \label{eq:Edip2}
\end{equation}

First, we used Eq.~\ref{eq:Edip} to calculate the shape anisotropy energy of the undoped \vte{}. Then, we used Eq~\ref{eq:Edip2} to calculate the $E_{SA}$ of \vse{} and \vs{} at different doping levels. The shape anisotropy energy values from the dipolar energy in both pristine and charge-doped \vx{} are summarized in Table~\ref{tab-a:SA}. The magnitude of the shape anisotropy energy is less than 20~$\mu$eV in all three representative \vx{} systems within the doping levels that we considered.

\begin{table}[h!]
\centering
\caption{Shape anisotropy energy [in $\mu$eV] in \vx{} at different doping levels $\delta$ per cell.}
\begin{tabular}{l l l l}\toprule
$\delta$ & \vte{} & \vse{} & \vs{} \\\midrule
-0.10 & -16.1 & -14.1 & -11.7 \\
-0.05 & -15.5 & -14.8 & -12.3 \\
0 & -15.0 & -14.4 & -13.1 \\
0.05 & -15.9 & -13.9 & -12.5 \\
0.10 & -16.7 & -13.3 & -11.9 \\ \bottomrule
\end{tabular}
\label{tab-a:SA}
\end{table}

\newpage
\section{Projected bands for \vte{} with $U=1.3$~eV}
\label{section:proj_bands_U1.3}

The contribution of each V $d$ and Te $p$ orbital in \vte{} is shown in Figures \ref{fig:fat_all_V_u1.3} and \ref{fig:fat_all_Te_u1.3}, respectively. The VBM at $\Gamma$ has $d_{xz}$/$d_{yz}$ character from V and $p_x$/$p_y$ character from Te.

\begin{figure}[h]
    \centering
    \includegraphics[width=1\textwidth]{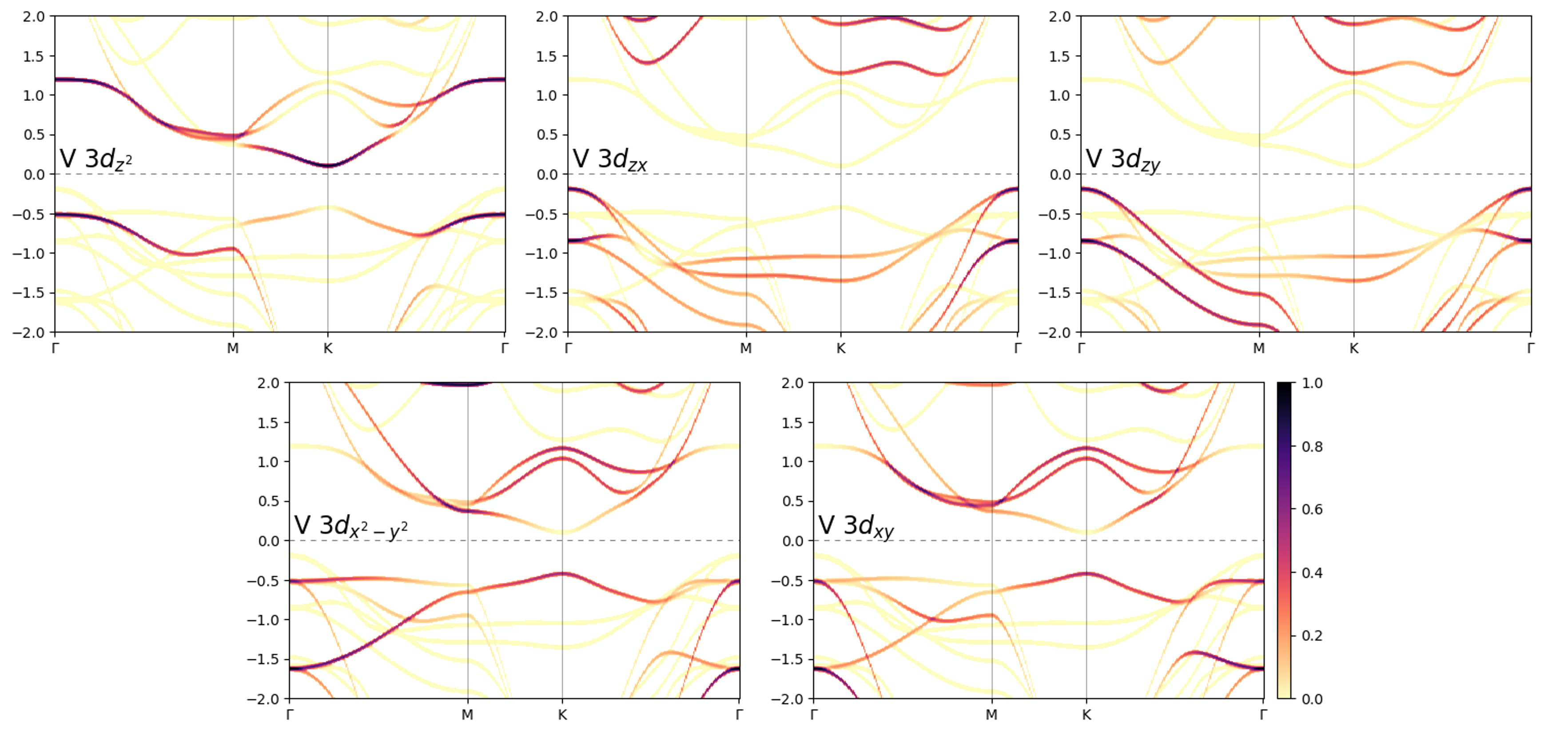}
    \caption{Band structure with orbital-projected contributions from the V atom in \vte{}.The color scale and the bar on the right indicate the projection strength of each orbital/state. The Fermi level is set to zero.}
    \label{fig:fat_all_V_u1.3}
\end{figure}

\begin{figure}[h]
    \centering
    \includegraphics[width=1\textwidth]{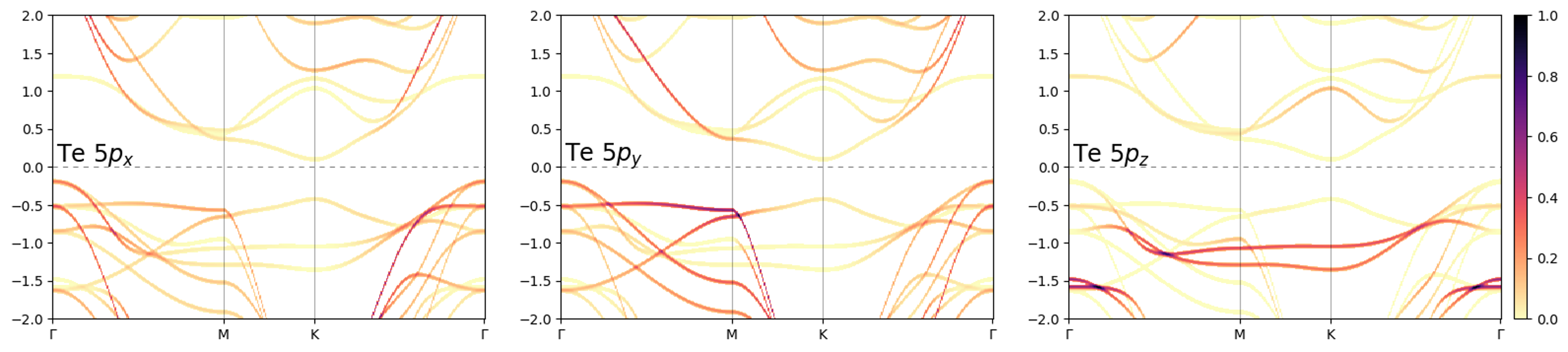}
    \caption{Band structure with orbital-projected contributions from the Te atom in \vte{}. The color scale and the bar on the right indicate the projection strength of each orbital/state. The Fermi level is set to zero.}
    \label{fig:fat_all_Te_u1.3}
\end{figure}

\newpage

\section{Doping-dependent  magnetic anisotropy model: VBM splitting and emptied states integration}
\label{section:empty_doping}

We develop a simple model to estimate the energy difference between OOP and IP orientations as a function of hole doping concentration, $\delta$, within the rigid band approximation. The total energy difference is expanded to second order in $\delta$:
\begin{equation}
\Delta E_{\rm{total}} = A \delta + B \delta^2.
\end{equation}

The linear coefficient $A = \epsilon^{\rm{VBM}}_{\rm{OOP}} - \epsilon^{\rm{VBM}}_{\rm{IP}}$ represents the energy offset at the valence band maximum for each doping level. This term dominates at low doping levels, where holes occupy states near the band edge. It is a direct consequence of the larger spin-orbit splitting of the OOP VBM. 

The quadratic coefficient $B$ captures how the energy difference evolves with increasing doping. Physically, it arises from the different band dispersions of the OOP and IP orientations. We evaluate this term by integrating the energy difference over the emptied states in the Brillouin zone:
\begin{equation}
B\delta^2 \approx \frac{1}{A_{BZ}} \int_{\rm{emptied\ states}} E_{\rm OOP}(\boldsymbol{k})d^2\boldsymbol{k} - \frac{1}{A_{BZ}}\int_{\rm{emptied\ states}}E_{\rm{IP}}(\boldsymbol{k}) d^2\boldsymbol{k},
\end{equation}
where $E_{\rm OOP,IP}(\boldsymbol k)$ denotes the energy function of $\boldsymbol k$ (without a constant term). 
The quadratic scaling emerges because both the integration area and the average energy difference within this volume grow linearly with $\delta$. This formulation naturally accounts for differences in effective mass and band curvature between the two orientations.

The complete expression thus becomes:
\begin{equation}
\label{eq:model}
     \Delta E_{\rm{total}}= \left[ \epsilon^{\rm{VBM}}_{\rm{OOP}} -\epsilon^{\rm{VBM}}_{\rm{IP}} \right ] \delta + \int_{\rm emptied\ states} E_{\rm OOP}(\boldsymbol{k}) d^2 \boldsymbol{k} - \int_{\rm emptied\ states} E_{\rm{IP}} (\boldsymbol{k}) d^2 \boldsymbol{k}. 
\end{equation}

The form of the energy dependence as a function of $\delta$ is particularly simple in the case of a single VBM peak in each of the OOP and IP cases. Particularly, if we approximate the peak with an isotropic paraboloid, the following can be written:
\begin{equation}
E_{\rm OP}(k) = \epsilon^{\rm VBM}_{\rm OP} - a_{\rm OP} k^2    
\end{equation}
and
\begin{equation}
E_{\rm IP}(k) = \epsilon^{\rm VBM}_{\rm IP} - a_{\rm IP} k^2.   
\end{equation}
For each case, the maximum wavevector $k_{\rm emptied}$ delimits the region of integration inside the paraboloid and 
is determined by:
\begin{equation}
k_{\rm emptied}= \frac{\delta A_{BZ}}{\pi},
\end{equation}
where $A_{BZ}$ denotes the area of the Brillouin zone.
Then the integral of the emptied states in the OOP case can be expressed as

\begin{align}
    \Delta E_{\rm OP} & = \frac{1}{A_{BZ}}\int_{\rm emptied\ states} (\epsilon^{\rm VBM}_{\rm OP} - a_{\rm OP} k^2) 2\pi k dk \\
     & \quad = \epsilon^{\rm VBM}_{\rm OP}\cdot \delta  - a_{\rm OP} \frac{A_{BZ}}{2\pi} \delta^2,
\end{align}

and similarly for $  \Delta E_{\rm IP}$. Finally, 
\begin{align}
\Delta E &=  \Delta E^0_{\rm OP} -  \Delta E^0_{\rm IP}\\
&= (\epsilon^{\rm VBM}_{\rm OP} - \epsilon^{\rm VBM}_{\rm IP}) \cdot \delta - \frac{A_{BZ}}{2\pi}(a_{\rm OP} - a_{\rm IP})\cdot \delta^2.
\label{eq:two_parabolas}
\end{align}
 Equation \ref{eq:two_parabolas} shows that the quadratic term's coefficient is determined by the difference of effective masses of the individual peaks.

In the case of \vte{}, we have two peaks in the IP orientation and we cannot directly apply the Eq.~\ref{eq:two_parabolas}. Instead, we used the numerical integration of the fitted paraboloids as in Eq.~\ref{eq:model} and we obtained the energy differences presented in  Fig~\ref{fig:Eempt_MCA_doping}. The data reveal that both $\Delta E$ and the MCA increase linearly with small hole doping. A minor deviation is observed for moderate hole doping (around $\delta>+0.02$~h/cell).

\begin{figure}[h!]
    \centering
    \includegraphics[width=0.5\linewidth]{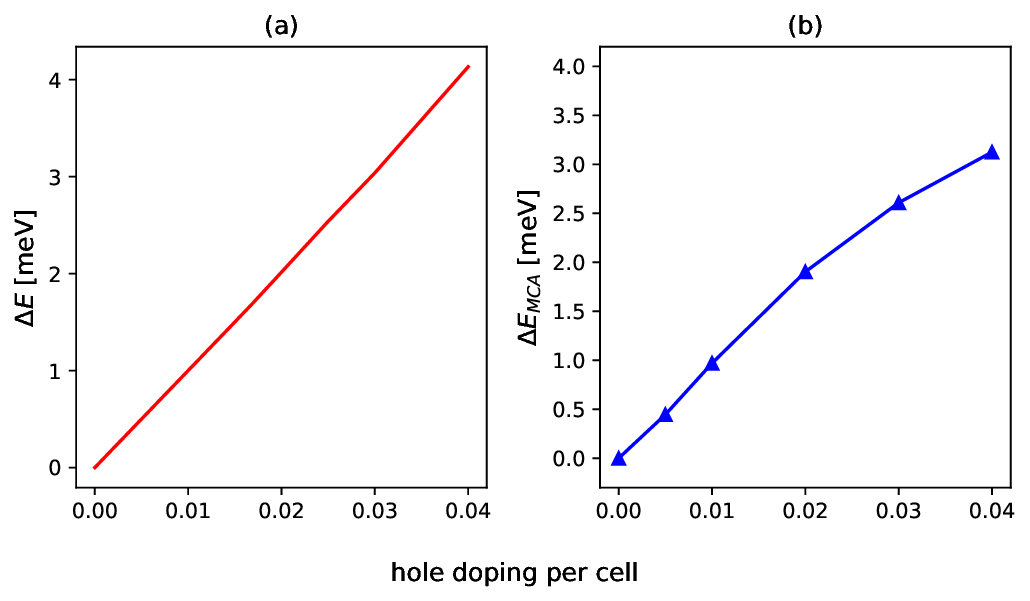}
    \caption{(a) Difference of the energies of the VBM in the OOP and IP magnetization directions per hole doping level, and the (b) MCA relative to the undoped per hole doping level in \hvte{}. }
    \label{fig:Eempt_MCA_doping}
\end{figure}

\clearpage
\newpage
\section{Projected bands for \vs{} and \vse{} with $U=1.3$~eV}
\label{section:proj_bands_U1.3_VSe_VS}

In both \vs{} and \vse{}, the highest valence state at $\Gamma$ is dominated by V $d_{z^2}$ while the one at K is dominated by V $d_{xy}$/$d_{x^2-y^2}$ orbitals.

The contribution of each V $d$ and Se $p$ orbital in \vse{} bands is shown in Figures \ref{fig:VSe2_fat_all_V_u1.3} and \ref{fig:VSe2_fat_all_Se_u1.3}, respectively.

\begin{figure}[h]
    \centering
    \includegraphics[width=1\textwidth]{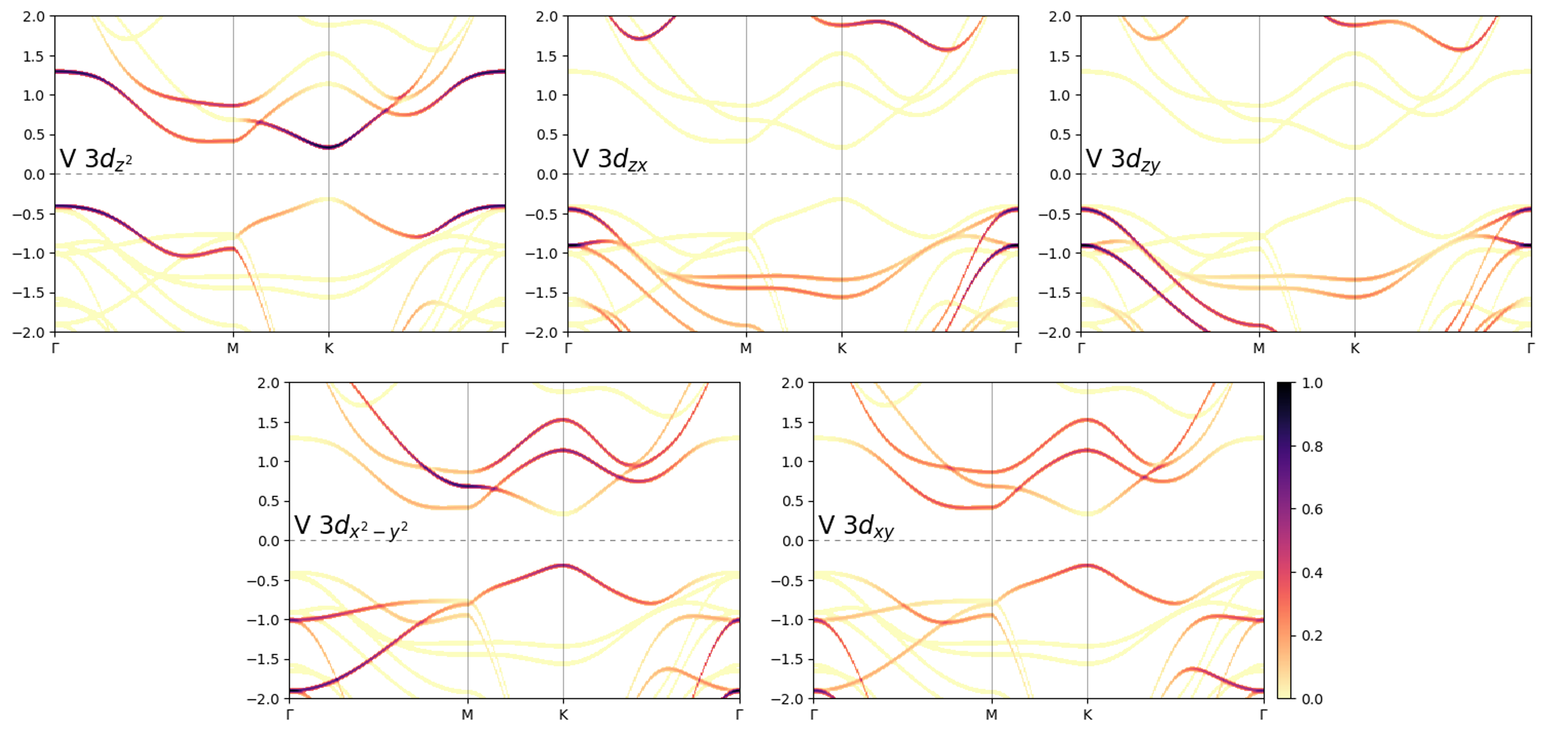}
    \caption{Band structure with orbital-projected contributions from the V atom in \vse{}. The color scale and the bar on the right indicate the projection strength of each orbital/state. The Fermi level is set to zero.}
    \label{fig:VSe2_fat_all_V_u1.3}
\end{figure}

\begin{figure}[h]
    \centering
    \includegraphics[width=1\textwidth]{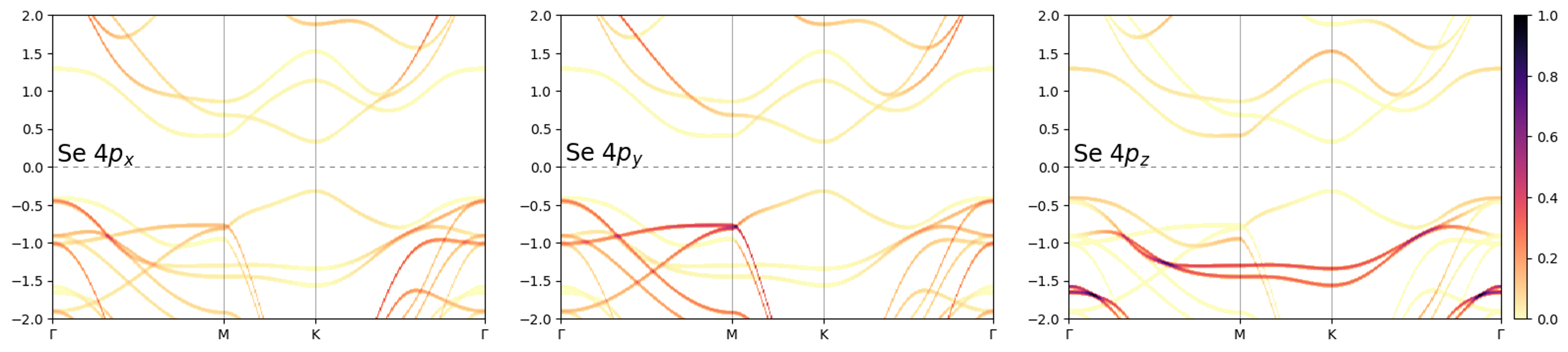}
    \caption{Band structure with orbital-projected contributions from the Se atom in \vse{}. The color scale and the bar on the right indicate the projection strength of each orbital/state. The Fermi level is set to zero.}
    \label{fig:VSe2_fat_all_Se_u1.3}
\end{figure}

\newpage
The contribution of each V $d$ and S $p$ orbital in \vs{} is shown in Figures \ref{fig:VS2_fat_all_V_u1.3} and \ref{fig:VS2_fat_all_S_u1.3}, respectively.

\begin{figure}[h]
    \centering
    \includegraphics[width=1\textwidth]{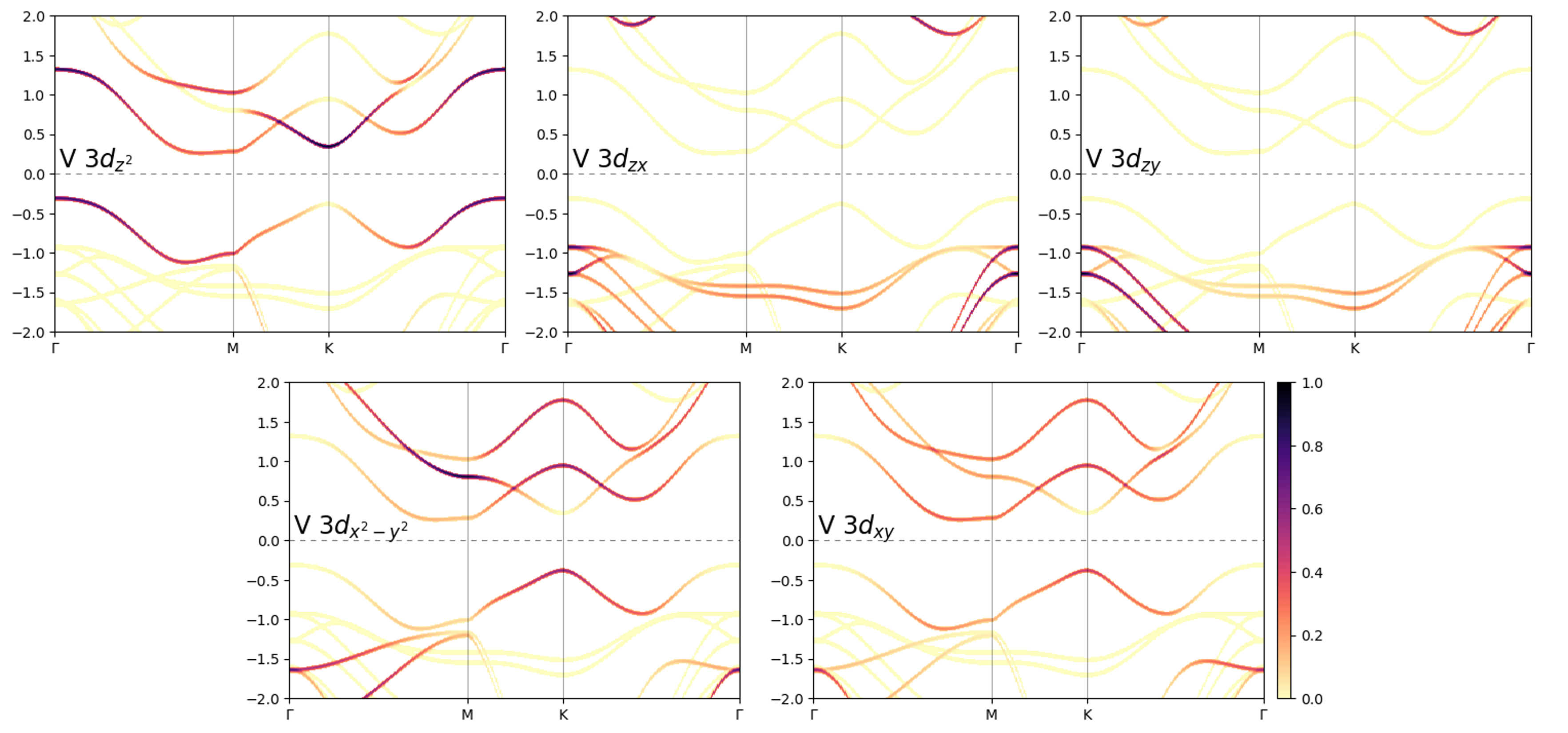}
    \caption{Band structure with orbital-projected contributions from the V atom in \vs{}. The color scale and the bar on the right indicate the projection strength of each orbital/state. The Fermi level is set to zero.}
    \label{fig:VS2_fat_all_V_u1.3}
\end{figure}

\begin{figure}[h]
    \centering
    \includegraphics[width=1\textwidth]{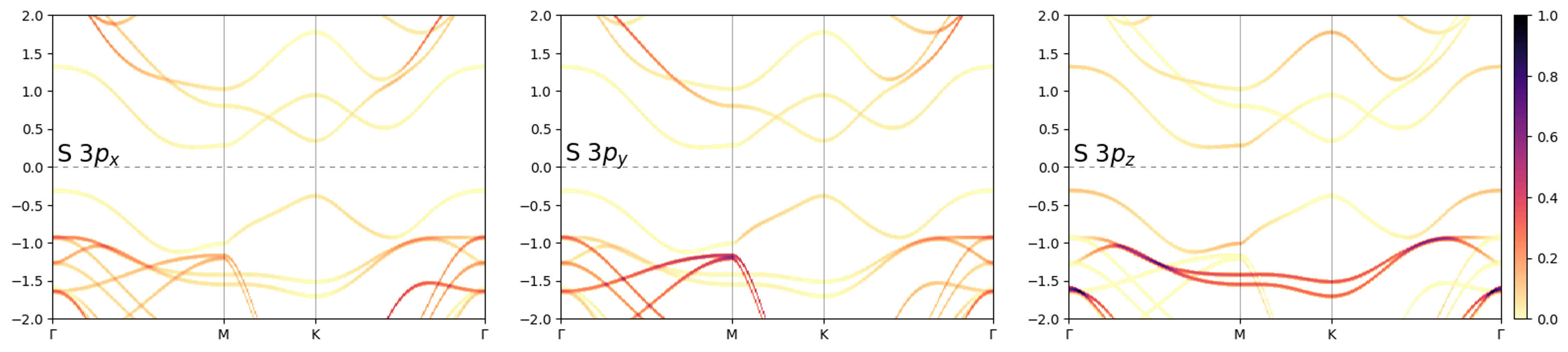}
    \caption{Band structure with orbital-projected contributions from the S atom in \vs{}. The color scale and the bar on the right indicate the projection strength of each orbital/state. The Fermi level is set to zero.}
    \label{fig:VS2_fat_all_S_u1.3}
\end{figure}

\clearpage
\newpage
\section{Valley splitting}
\label{section:valley}

It has been shown that the \hvx{} monolayers have an intrinsic valley splitting due to SOC and the absence of inversion symmetry.  However, the mechanism of switching to PMA by hole doping is unaffected by valley splitting which can be found, for example, at the K$_{\pm}$ points given by $\vec{k}=(\pm\frac{2}{3},0,0)\frac{\pi}{2a}$.
In \vte{}, the VBM arises from degenerate d orbitals at $\Gamma$, so the PMA is not influenced by the valley splitting at K$_{\pm}$.  In contrast, for \vse{} and \vs{}, the spin–orbit shift of the highest valence state at K$_{\pm}$ becomes relevant: in \vse{} this state forms the VBM, and in \vs{} this SO-shifted peak can significantly affect the magnetic anisotropy, even if it is not the VBM. Here we will show that the energy of the states emptied when the magnetization is OOP is still higher than when it is IP even when taking into account valley polarization. We demonstrate this below for \vse{}, and the analysis can be extended  to the valleys of \vs{}. 

Plotting the fully relativistic bands of \vse{} along the $\Gamma$---K$_{-}$---M---K$_{+}$---$\Gamma$ path in the BZ as in Figure \ref{fig-a:VSe2-valley}, we observe a valley splitting between the highest valence states at the K$_{\pm}$ points of $\Delta E_{K\pm}^{\perp}=109$~meV in the OOP case for the pristine \vse{}. The valley splitting is negligible in the IP case for both doping levels.

\begin{figure}[h!]
    \centering
    \includegraphics[width=.45\linewidth]{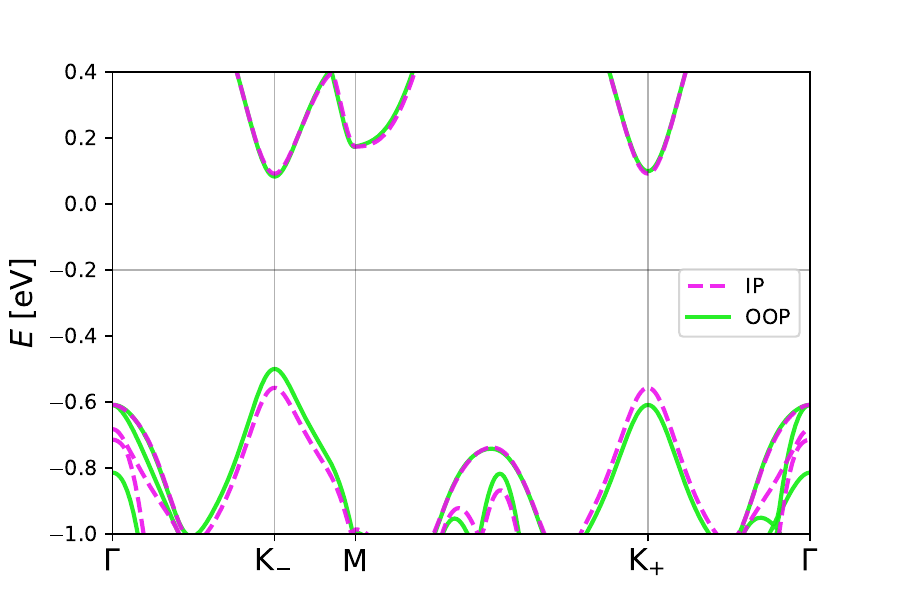}
    \includegraphics[width=.45\linewidth]{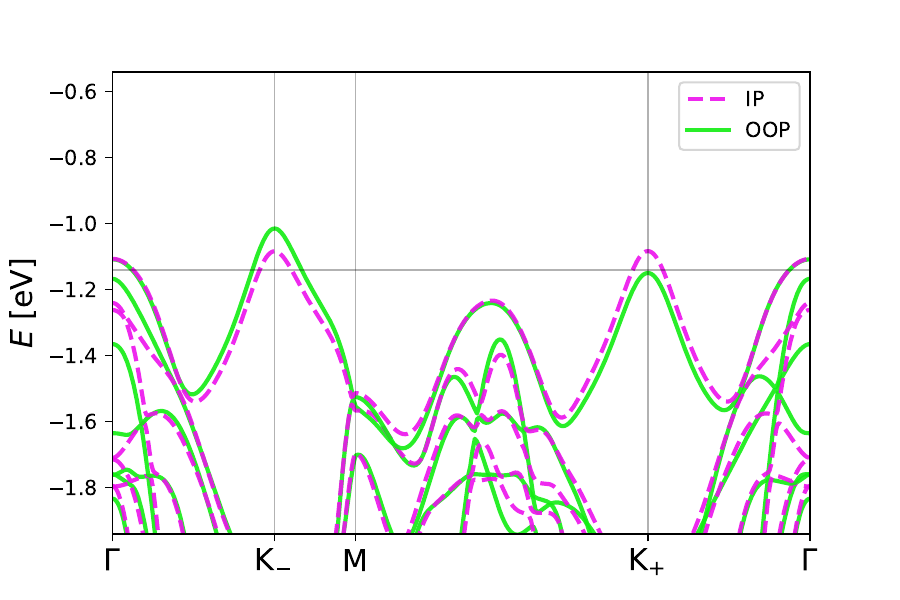}
    \caption{Fully relativistic band structures with IP and OOP magnetization of the undoped (left) and $\delta=+0.05$ hole-doped (right) \vse{}. The bands are aligned with respect to the respective vacuum potentials. The horizontal grey lines denote the Fermi level, which at the hole doping of 0.05 is similar for the OOP and IP cases---the difference is only 8~meV.}
    \label{fig-a:VSe2-valley}
\end{figure}

Based on our analysis from SI Section~\ref{section:empty_doping}, we can write expressions for the contribution to the MAE originating from emptying of the bands in the valley split VBM. To simplify the analysis, we will assume that both the IP and OOP peaks at  $K_\pm$ can be described as second-order paraboloids with the same curvature. This is a good approximation for the VBM peaks at K$_\pm$, as the main difference between the peaks in the OOP and IP orientation is the shift in energy. Since \vse{} exhibits virtually  no valley splitting for the IP magnetic orientation, we  considering valley splitting only in the OOP orientation,  which results in two distinct paraboloids at $K_+$ and $K_-$:
\begin{equation}
E_{\rm OP,K+} = \epsilon^{\rm VBM}_{\rm OP,K+} - a k^2    
\end{equation}
and
\begin{equation}
E_{\rm OP,K-} = \epsilon^{\rm VBM}_{\rm OP,K-} - a k^2.    
\end{equation}

We also disregard the presence of the peak at $\Gamma$, which is identical in the IP and OOP cases, so its main effect is  shifting the critical hole doping concentrations to a larger one.  We consider two cases: (i) when the highest OP and two degenerate IP  peaks are being depleted and (ii) when both OP peaks are being emptied (together with two degenerate IP peaks). 

\vspace{1cm}
{\bf{Case(i): one OP peak emptied}}
\vspace{3mm}

In this case, the  effective doping in the IP paraboloids will be exactly half of the doping in the OP case, but simultaneously, there will be twice the contribution from the IP due to the two peaks. This yields the following expression:
\begin{equation}
\Delta E = (\epsilon^{\rm VBM}_{\rm OP,K+} - \epsilon^{\rm VBM}_{\rm IP}) \cdot \delta - \frac{A_{BZ}}{4\pi}a\cdot \delta^2.
\end{equation}
Let's consider the case when the Fermi level  in the OOP case is just above the lower band at the energy $\epsilon^{\rm VBM} _{\rm IP, K-}$. It determines the maximal k extent of the OOP paraboloid at $K_+$ in the three-band scenario as:
\begin{equation}
k^2_{\rm emptied} = \frac{\Delta_{\rm valley}}{a},
\end{equation}
where $\Delta_{\rm valley}$ denotes the valley splitting in the OOP case, $\Delta_{\rm valley} =\epsilon^{\rm VBM}_{\rm OP,K+}-\epsilon^{\rm VBM}_{\rm OP,K-}$. From there the largest doping still depleting only one OP and two IP bands is:
\begin{equation}
\delta = \frac{\Delta_{\rm valley}}{a A_{BZ}}.
\end{equation}
Thus, $\Delta E$ at this doping is:
\begin{equation}
\Delta E = \frac{\Delta_{\rm valley}}{a A_{BZ}}\left[\epsilon^{\rm VBM}_{\rm OP} - \epsilon^{\rm VBM}_{\rm IP} - \frac{\Delta_{\rm valley}}{4\pi}\right]. 
\label{eq:three_bands}
\end{equation}

\vspace{1cm}
{\bf{Case(ii): two OP peaks emptied}}
\vspace{3mm}

We now focus on the situation in which  the peaks at all $K_\pm$ are partially emptied (two in OOP and two in IP orientations). In the OOP orientation, the two peaks at different energies are depleted up to a common Fermi level:
\begin{equation}
E_{\rm F,  OP}=\frac{\epsilon^{\rm VBM}_{\rm OP,K+}+\epsilon^{\rm VBM}_{\rm OP,K-}}{2} -\frac{a A_{BZ}}{2\pi}\delta. 
\end{equation}
and, consequently, with different effective doping levels for each peak, such that $\delta_{K_+} + \delta_{K_-} = \delta$:
\begin{equation}
\delta_{K+} = \frac{\delta + \frac{\pi \, \Delta_{\rm valley}}{a \, A_{\rm BZ}}}{2}
\end{equation}
and
\begin{equation}
\delta_{K-} = \frac{\delta - \frac{\pi \, \Delta_{\rm valley}}{a \, A_{\rm BZ}}}{2}.
\end{equation}
 Finally, the energy difference between the OOP and IP orientations in this case can be written as:
\begin{equation}
\Delta E = \frac{\pi  \Delta^2_{\rm valley}}{4 a A_{BZ}} + \left[\frac{\epsilon^{\rm VBM}_{\rm OP,K+}+\epsilon^{\rm VBM}_{\rm OP,K-}}{2}  - \epsilon^{\rm VBM}_{\rm IP}  \right]\delta.
\label{eq:four_bands}
\end{equation}
It consists of a constant term depending on the OOP valley splitting and a linear term which is positive if the IP VBM is below the average of the valley-split OOP peaks. 

In \vse{},  $\Delta_{\rm valley} = 109$~meV in the OOP orientation and only 0.4~meV in the IP orientation. When aligned to the vacuum potential, the difference $\epsilon^{\rm VBM}_{\rm OP,K+} - \epsilon^{\rm VBM}_{\rm IP} = 57$~meV and $\epsilon^{\rm VBM}_{\rm OP,K-} - \epsilon^{\rm VBM}_{\rm IP} = -52$~meV. This clearly means that $\Delta E$ from Eq.~\ref{eq:three_bands}  and Eq.~\ref{eq:four_bands} is positive.

\clearpage
\newpage

\newpage

\clearpage
\newpage
\section{Trends in \vx{}}
\label{section:trends_vx2}

The energy splittings and shifts mentioned in the main text are summarized in Table \ref{tab-a:split_shift_summ}. The crystal field splitting is calculated as the scalar-relativistic energy difference of two valence states in the same spin channel but with different orbital characters. The exchange splitting is calculated as the scalar-relativistic energy difference of two valence states with the same orbital character but different spin. The energy shift due to SOC is calculated as the energy difference of the fully relativistic state and the corresponding scalar-relativistic state. There is an increasing trend in crystal field splitting and a decreasing trend in exchange splitting and in spin-orbit effect from \vte{} to \vse{} to \vs{}. 

\begin{table}[h!]
\centering
\caption{Energy splittings and shifts due to crystal field (CF) effect, exchange (Ex) interaction, and spin-orbit (SO) coupling in \hvx{} ($X=$Te, Se, and S). The states are labeled as follows, A': $d_{z^2}$, E': $d_{xy}$ \& $d_{x^2-y^2}$, and E": $d_{xz}$ \& $d_{yz}$. Values are in eV.}
\begin{tabular}{@{}llll@{}}
\toprule
 & \vte{} & \vse{} & \vs{} \\ \midrule
CF splitting between A’ and E’’ states at $\Gamma$ & 0.333 & 0.501 & 0.761 \\
CF splitting between E’ and E’’ states at $\Gamma$ & 0.325 & 0.562 & 0.636 \\
Ex splitting between E' states at $\Gamma$ & 1.103 & 0.894 & 0.784 \\
Ex splitting between E'' states at $\Gamma$ & 0.655 & 0.459 & 0.350 \\
SO splitting between highest E'' states (IP) at $\Gamma$ & 0.018 & 0.032 & 0.026 \\
SO splitting between highest E'' states (OOP) at $\Gamma$ & 0.337 & 0.206 & 0.054 \\
SO shift in the highest E’ state (IP) at K & 0.025 & 0.007 & 0.001 \\
SO shift in the highest E’ state (OOP) at K & 0.124 & 0.064 & 0.036 \\ \bottomrule
\end{tabular}
\label{tab-a:split_shift_summ}
\end{table}

\subsection{Hybridization at $\Gamma$}

The coefficients $|\langle \phi_{n l m} | \psi_{n k} \rangle|^2$ from the projections of the bands at $\Gamma$ onto the atomic orbitals are shown in Figure \ref{fig-a:trends-hybrid}. There is an increasing V-$X$ hybridization from \vte{} to \vs{}.

\begin{figure}[h!]
    \centering
    \includegraphics[width=\linewidth]{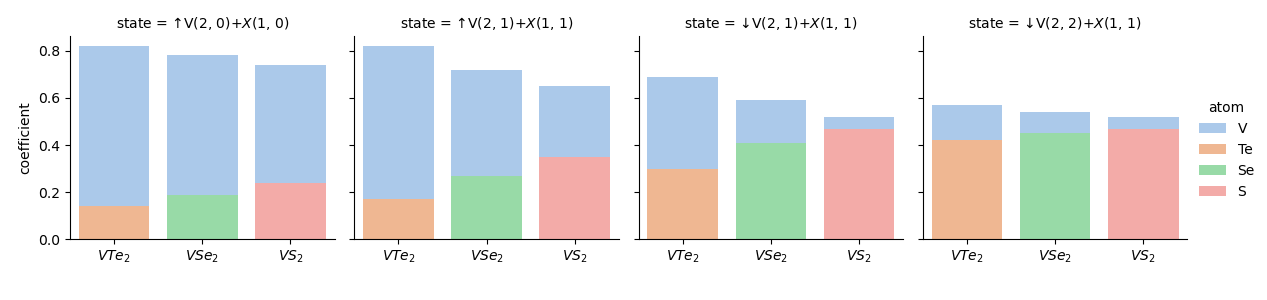}
    \caption{Projection onto atomic states at $\Gamma$ in \vx{} ($X=$ Te, Se, and S). The states are labeled by the spin, atom symbol, and the $(l,m_l)$ quantum numbers of the orbitals.}
    \label{fig-a:trends-hybrid}
\end{figure}

\newpage
\subsection{Partial density of states}

The PDOS plots in Figure \ref{fig-a:PDOS_sp} show an increasing hybridization from \vte{} to \vs{}. Among the \vx{} structures, \vte{} has the most localized states.

\begin{figure}[h!]
    \centering
    \includegraphics[width=.5\linewidth]{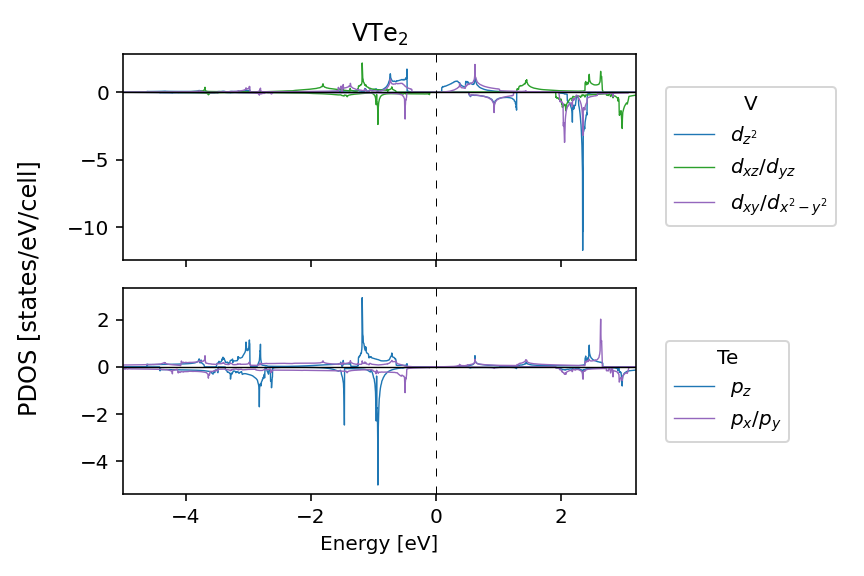}
    \includegraphics[width=.5\linewidth]{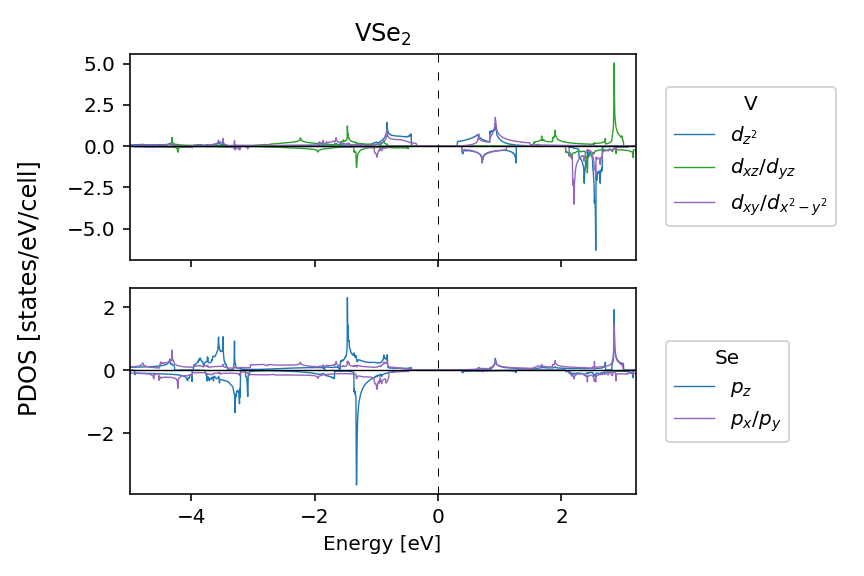}
    \includegraphics[width=.5\linewidth]{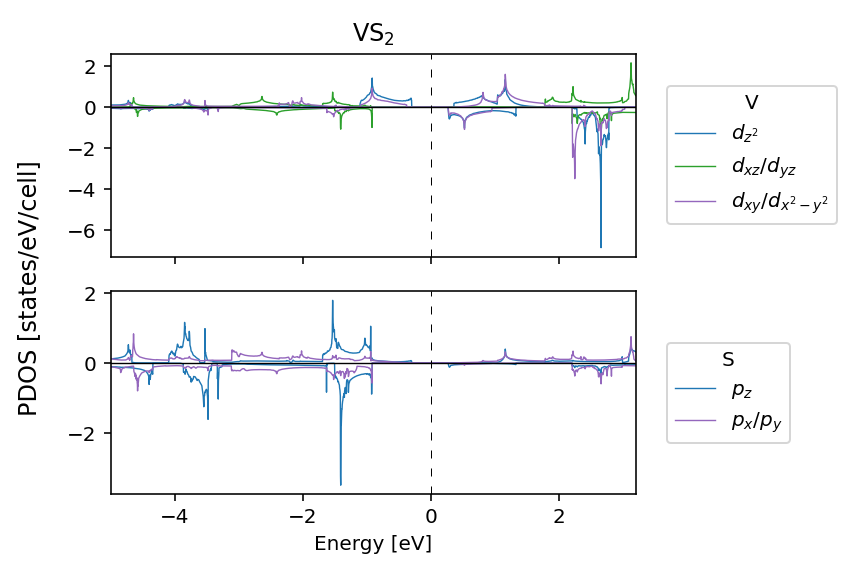}
    \caption{Partial density of states projected onto the vanadium $d$ and chalcogen $p$ spin-orbitals of \vte{} (top), \vse{} (middle) and \vs{} (bottom). The positive/negative PDOS corresponds to majority/minority spin states.}
    \label{fig-a:PDOS_sp}
\end{figure}

\newpage
\subsection{V magnetic moments}

The increasing hybridization and decreasing exchange splitting from \vte{} to \vs{} imply a trend of decreasing magnetic moment $m$ of vanadium in \vx{}. For instance, the vanadium magnetic moments in the pristine structures are as follows: $m_{VTe_2}=1.25\mu_B>m_{VSe_2}=1.07\mu_B > m_{VS_2}= 0.96\mu_B$.

The effect of charge doping on the magnetic moment can be predicted from the spin-polarized bands (Figures \ref{fig:sp_bands_u1.3}a and \ref{fig:mca_vx2}) of pristine \vx{} calculated at the scalar-relativistic level. The presence of majority/minority spins in the valence band edge will contribute to the decrease/increase in magnetic moment upon hole doping. Fig.~\ref{fig:uB_VX2} shows the V magnetic moment at different doping levels for the three monolayers. 
In \vte{}, the valence band edge is in the minority spin channel. Emptying exclusively these states increases $m$.
On the other hand, the depletion of majority-spin valence band upon hole doping in both \vse{} and \vs{} makes $m$ smaller. Similar arguments can be made for electron doping except that, instead, the presence of majority/minority spins in the conduction band edge will contribute to the increase/decrease in $m$.

\begin{figure}[h!]
    \centering
    \includegraphics[width=1\linewidth]{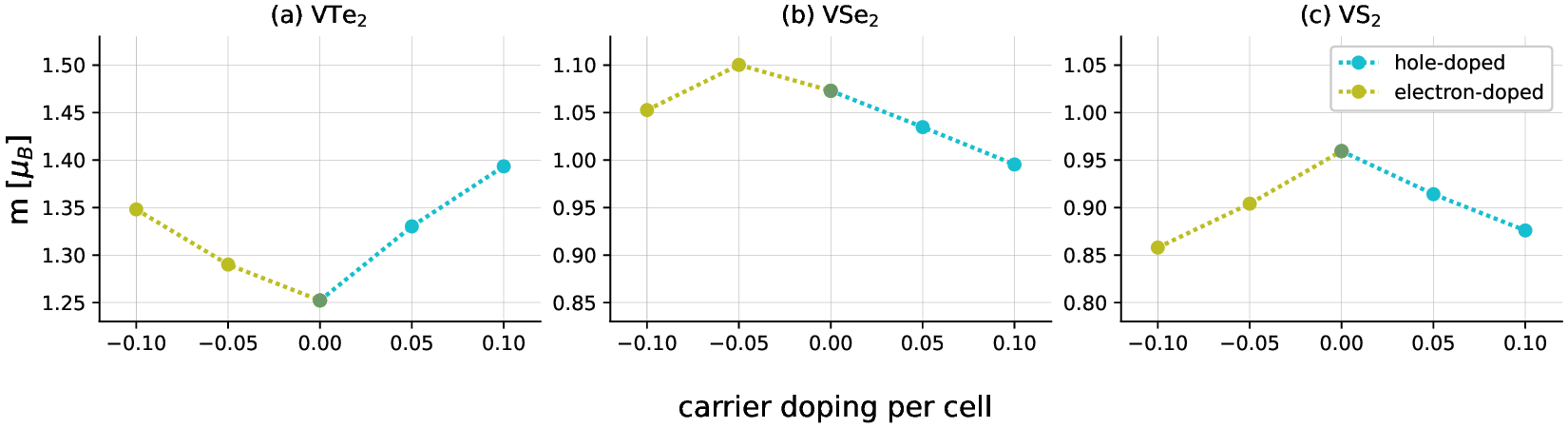}
    \caption{Absolute value of vanadium magnetic moment versus charge carrier doping concentrations $\delta$ for (a) \vte{}, (b) \vse{}, and (c) \vs{}. Hole doping corresponds to $\delta>0$ while electron doping $\delta<0$.}
    \label{fig:uB_VX2}
\end{figure}

In general, changes in the magnetic moment upon weak electron or hole doping can be predicted well from the spin-polarized scalar-relativistic bands if the conduction or valence band edges are dominated by a single spin channel, as in the case of \hvx{}.
This also renders all three \hvx{} monolayers half-metallic upon minimal charge doping. We also note that weak charge doping---within the doping levels that we considered in Fig.~\ref{fig:uB_VX2}---has been shown to maintain the ferromagnetic ordering of \hvx{} systems \cite{tang2022strain,jiang2024electronic,chen2024first}.

\newpage
\section{Comparisons at different $U$}
\label{section:comp_diffU}

We implemented GGA+$U$ calculations in Quantum Espresso using L\"owdin orthogonalized atomic orbitals to build the Hubbard projectors.
The inclusion of Hubbard $U$ can change the electronic and magnetic properties of materials. In \vte{}, increasing $U$ pushes the minority-spin conduction bands and the majority-spin valence bands away from the Fermi level: see insets in Fig.~\ref{fig:mca_all_with_bands}. The main changes consist in
lowering the majority-spin valence bands while raising the minority-spin band with the strongest effect occurring at the K point.
In lifting the minority-spin CBM from its intermediate $\Gamma$---K position in the pure GGA calculation, the CBM is shifted to a majority-spin state at K with GGA+$U$; on the other hand, the degenerate minority-spin VBM remains at $\Gamma$.
Despite these changes in the electronic bands, Fig.~\ref{fig:mca_all_with_bands} shows that the switching to PMA upon hole doping occurs regardless of the $U$ parameter: ${\rm MAE}>0$ at $\delta=+0.05$~h/cell for $U=$ 0, 1.3, and 1.7 eV. This can be attributed to the doubly degenerate VBM states at $\Gamma$ that persist even with increasing $U$.

\begin{figure}[h!]    \centering
    \includegraphics[width=1\textwidth]{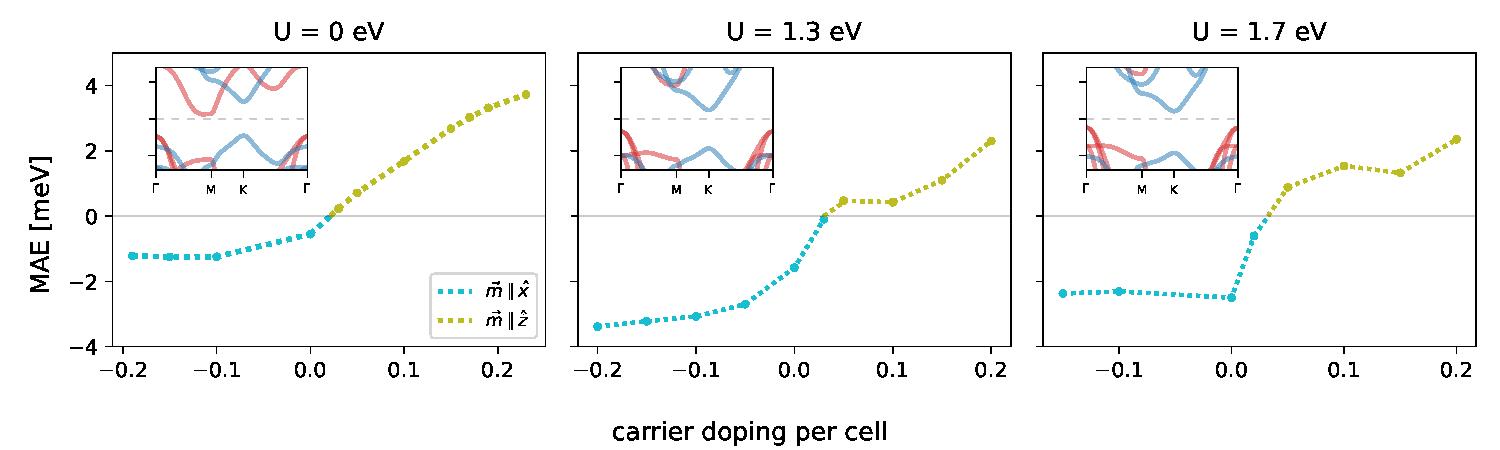}
    \caption{MAE vs carrier doping $\delta$ for different values of $U$ in \hvte{}. The insets show the bands close to the Fermi level; blue/red curves correspond to majority/minority spin bands.}
    \label{fig:mca_all_with_bands}
\end{figure}

The MAE of pristine \hvte{} (-1.58~meV) with $U=1.3$~eV is comparable to those in the literature employing GGA+U with values ranging from -1.5 to -1.9~meV, all favoring easy in-plane magnetization in the pristine structure. The calculated band gap ($E_g=0.29$~eV in our study vs. literature values ranging from 0.16 to 0.38 eV) and the VBM (minority-spin bands at $\Gamma$) are also similar to those in the literature \cite{fuh2016newtype, jafari2023electronic, tang2022strain, chen2020electronic, wang2021effects}.

Figure \ref{fig:sp_all} shows how $U$ affects the energy bands of the pristine \hvte{}. 
Increasing $U$ raises the lowest minority-spin conduction band and the highest minority-spin valence band around $\Gamma$ while attracting the majority-spin states towards the Fermi level at K, thus decreasing the electronic band gap.
These effects are much stronger in $U=2.0$~eV and $U=2.5$~eV where the system becomes a metal.
For smaller $U$ values, the system is a semiconductor with the VBM at $\Gamma$ and the CBM at M for $U=0$ and at K for both $U=1.3$~eV and $U=1.7$~eV; the band gaps are 0.30~eV without $U$, 0.29~eV with $U=1.3$~eV and 0.22~eV with $U=1.7$~eV. 

\begin{figure}[h!]
    \centering
    \includegraphics[width=1\textwidth]{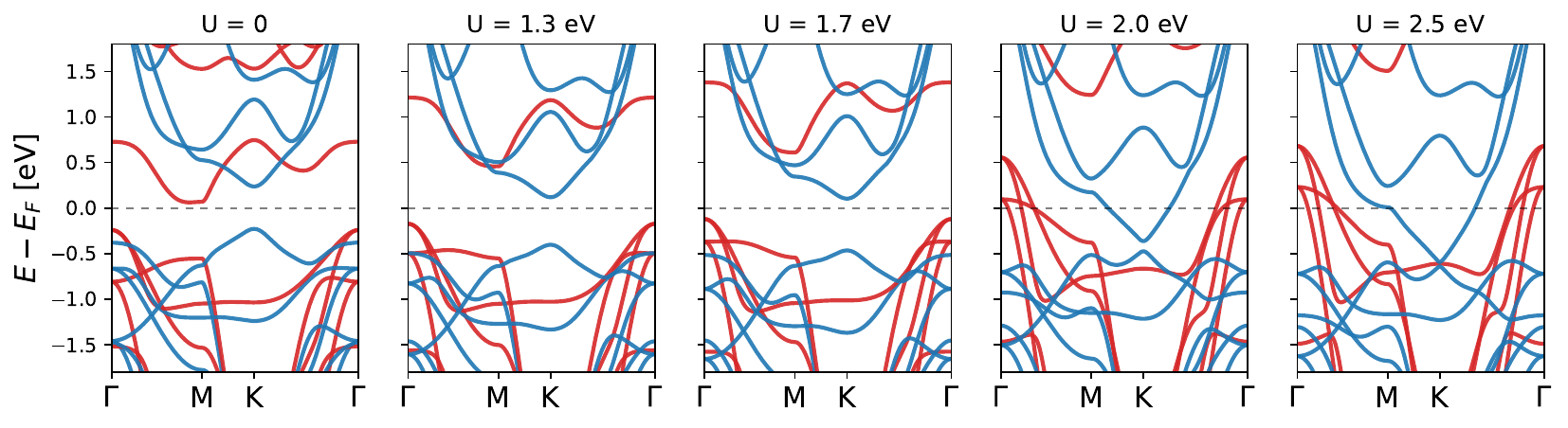}
    \caption{Spin-polarized band structures  of \vte{} calculated with increasing $U$ parameter. The blue/red bands represent majority/minority spin channels. The Fermi level is set to zero.}
    \label{fig:sp_all}
\end{figure}

\newpage
Figure \ref{fig:uB_all} shows the vanadium magnetic moment $m$ in \vte{}  at different doping levels with different U values. The magnetic moment without U decreases with larger carrier doping concentrations at the GGA level, while it increases as $|\delta|$ increases at the GGA+$U$ level for both $U=1.3$~eV and $U=1.7$~eV. Since $m$ depends on the number of occupied majority- and minority-spin states, and $U$ alters the conduction and valence bands near $E_F$ (cf. for example CBM without $U$ and with $U=1.3$~eV in Fig.~11 in the main text), the choice of $U$ influences the change in $m$  based on the spin character of the occupied/emptied states upon electron/hole doping. Therefore, the magnetic moment---a property that does not originate from SOC---is affected by the Hubbard U parameter.

\begin{figure}[h!]
    \centering
    \includegraphics[width=1\textwidth]{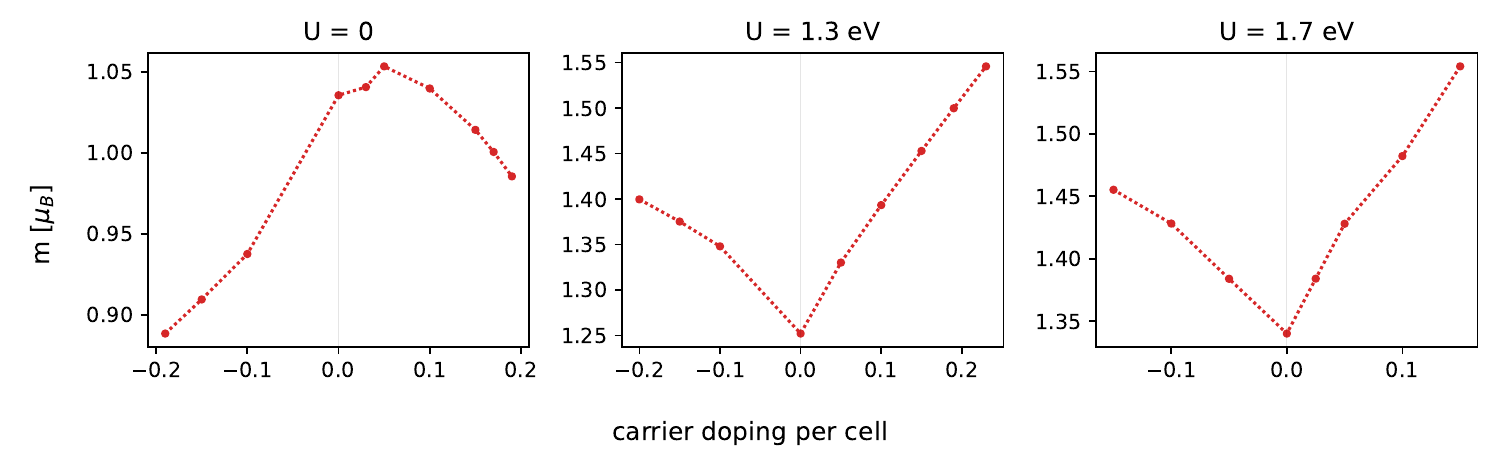}
    \caption{The vanadium magnetic moment  vs carrier doping $\delta$ for different values of $U$ in \vte{}. Hole doping corresponds to a positive value on the x-axis while electron doping is negative.}
    \label{fig:uB_all}
\end{figure}

For the MAE of \hvse{} (-0.70~meV) and \hvs{} (-0.17~meV), the MAE values at $U=1.3$~eV both fall within the previously reported range of -0.58 to -0.71~meV for the former \cite{jiang2024electronic,fuh2016newtype, jafari2023electronic} and -0.14 to -0.22~meV for the latter \cite{chen2024first,fuh2016newtype, jafari2023electronic}, all favoring easy in-plane magnetization in the pristine structure.  The band gaps ($E_g=0.65$~eV for \hvse{} and $E_g=0.57$~eV for \hvs{}) and the VBM (majority-spin at K for \hvse{} and majority spin at $\Gamma$ for \hvs{}) are also similar to those in the literature ($E_g$ ranging from 0.55 to 0.68~eV for \hvse{} and from 0.35 to 0.72~eV for \hvs{}) \cite{jiang2024electronic, chen2024first, fuh2016newtype, jafari2023electronic, murila2021structural}. We note that for \hvs{}, the experimental band gap ranges from 0.1 to 1.55 eV \cite{su2021recent, joseph2023hydrothermally} while for \hvte{} and \hvse{}, no experimental measurement of the band gap has been performed to date.

Lastly, we note that in \vs{}---the system with the weakest SOC among the \vx{} monolayers---the value of $U$ does not significantly alter its MCA, e.g. in the pristine case, $E_{MCA}^{U=0}=-0.18$~meV, $E_{MCA}^{U=0.7}=-0.18$~meV, and $E_{MCA}^{U=1.3}=-0.17$~meV.

\clearpage
\newpage
\section{Point groups promoting increase in MCA upon hole doping}
\label{section:PG}

Here are the point groups (in Schoenflies notation) of materials that preserve degeneracies of given states based on crystallographic symmetry. These degeneracies, when found in the topmost valence bands, can promote PMA upon hole doping.

\begin{table}[h!]
\centering
\caption{Point groups that enforce degeneracies in both $d_{xz}$/$d_{yz}$ and $d_{x^2-y^2}$/$d_{xy}$ pairs, as well as $p_x$/$p_y$ degeneracies, are listed along with the irreducible representations (irreps) of these states.}
\begin{tabular}{l l l l}\toprule
 & $d_{xz}$/$d_{yz}$ & $d_{x^2-y^2}$/$d_{xy}$ & $p_x$/$p_y$ \\\midrule
$D_{6h}$ & E\textsubscript{1g} & E\textsubscript{2g} & E\textsubscript{1u} \\
$D_{3h}$ & E’’ & E’ & E’ \\
$C_{6v}$ & E\textsubscript{1} & E\textsubscript{2} & E\textsubscript{1} \\
$D_{6}$ & E\textsubscript{1} & E\textsubscript{2} & E\textsubscript{1} \\
$C_{6h}$ & E\textsubscript{1g} & E\textsubscript{2g} & E\textsubscript{1u} \\
$C_{3h}$ & E’’ & E’ & E’ \\
$C_{6}$ & E\textsubscript{1} & E\textsubscript{2} & E\textsubscript{1} \\
$D_{3d}$ & E\textsubscript{g} & E\textsubscript{g} & E\textsubscript{u} \\
$C_{3v}$ & E & E & E \\
$D_{3}$ & E & E & E \\
$C_{3i}$ & E\textsubscript{g} & E\textsubscript{g} & E\textsubscript{u} \\
$C_{3}$ & E & E & E \\ \bottomrule
\end{tabular}
\label{tab-a:pg1}
\end{table}

\begin{table}[h!]
\centering
\caption{Point groups that enforce degeneracies only in $d_{xz}$/$d_{yz}$ and $p_x$/$p_y$ pairs along with the irreducible representations (irreps) of these states.}
\begin{tabular}{l l l}\toprule
 & $d_{xz}$/$d_{yz}$ & $p_x$/$p_y$ \\\midrule
$O_{h}$ & T\textsubscript{2g} & T\textsubscript{1u} \\
$T_{d}$ & T\textsubscript{2} & T\textsubscript{2} \\
$O$ & T\textsubscript{2} & T\textsubscript{1} \\
$T_{h}$ & T\textsubscript{g} & T\textsubscript{u} \\
$T$ & T & T \\
$D_{4h}$ & E\textsubscript{g} & E\textsubscript{u} \\
$D_{2d}$ & E & E \\
$C_{4v}$ & E & E \\
$D_{4}$ & E & E \\ \bottomrule
\end{tabular}
\label{tab-a:pg2}
\end{table}

\clearpage
\newpage
\section{Mn$X_2$ ($X=$ I, Br, Cl)  }
\label{section:mnx2}

From the Computational 2D Materials Database \cite{haastrup2018computational}, we found a class of dynamically stable semiconducting monolayers Mn$X_2$ ($X=$ I, Br, Cl) which could exhibit enhanced PMA upon hole doping. These materials are ferromagnetic with an out-of-plane easy axis of magnetization. They have tetragonal coordination with $D_{2d}$ point group symmetry.

Considering MnBr\textsubscript{2} (https://c2db.fysik.dtu.dk/material/1MnBr2-2), we calculated an MCA of 0.2~meV in the pristine case in GGA. Upon $\delta=+0.05$ hole doping per cell, the MCA is increased to 1.5 meV. The increase in MCA can be attributed to the two-fold degenerate Mn $d_{xz}$/$d_{yz}$ orbitals hybridized with Br $p_x$/$p_y$ at the VBM at $\Gamma$ as shown by the orbital-projected bands in Fig.~\ref{fig:MnBr2_fat_all_Mn} and \ref{fig:MnBr2_fat_all_Br}. This secures a larger spin-orbit splitting at the VBM when $\vec{m}\parallel\hat{z}$ (180 meV) than when $\vec{m}\parallel\hat{x}$ (20 meV).

\begin{figure}[h]
    \centering
    \includegraphics[width=0.3\textwidth]{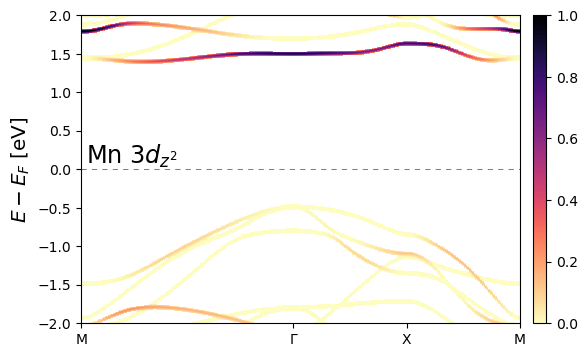}
    \includegraphics[width=0.3\textwidth]{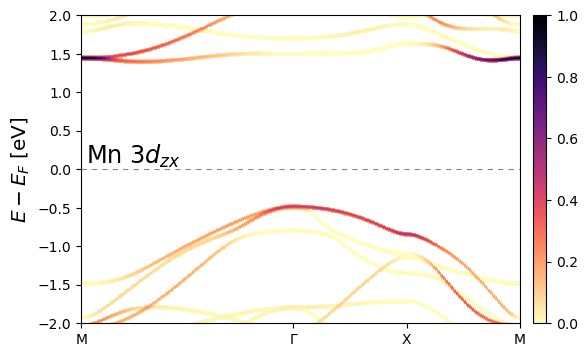}
    \includegraphics[width=0.3\textwidth]{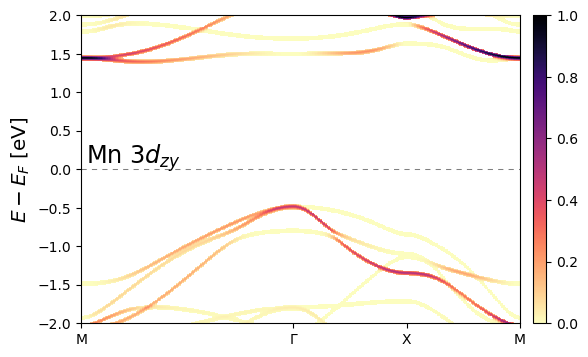}
    \includegraphics[width=0.3\textwidth]{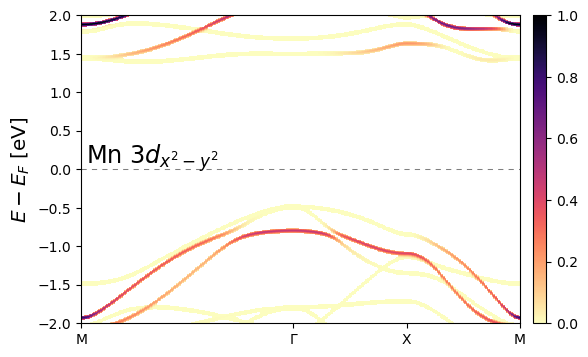}
    \includegraphics[width=0.3\textwidth]{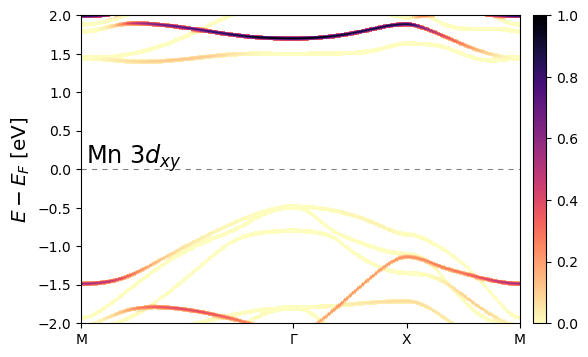}
    \caption{Band structure with orbital-projected contributions from the Mn $3d$ orbitals  in MnBr\textsubscript{2}. The color scale and the bar on the right indicate the projection strength of each orbital/state. The Fermi level is set to zero.}
    \label{fig:MnBr2_fat_all_Mn}
\end{figure}

\begin{figure}[h]
    \centering
    \includegraphics[width=0.3\textwidth]{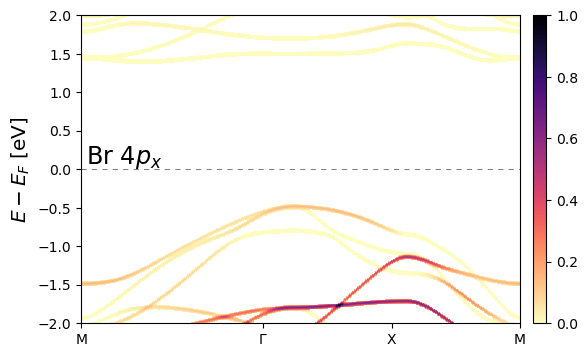}
    \includegraphics[width=0.3\textwidth]{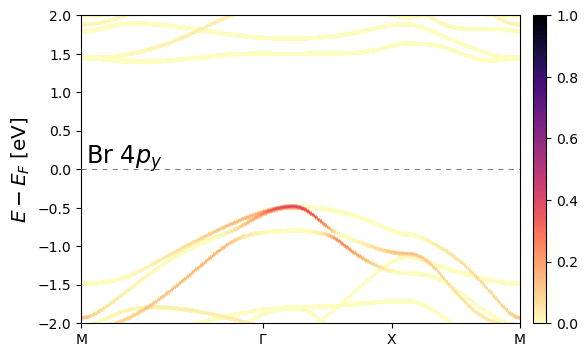}
    \includegraphics[width=0.3\textwidth]{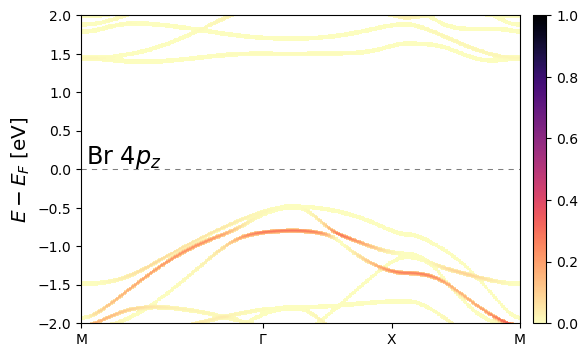}
    \caption{Band structure with orbital-projected contributions from the Br $4p$ orbitals in MnBr\textsubscript{2}. The color scale and the bar on the right indicate the projection strength of each orbital/state. The Fermi level is set to zero.}
    \label{fig:MnBr2_fat_all_Br}
\end{figure}

Since MnI\textsubscript{2} and MnCl\textsubscript{2} have similar band structures as MnBr\textsubscript{2}, we expect similar mechanism and behavior of the MCA upon hole doping in these materials.

\newpage

\endgroup

\putbib[references.bib]
\end{bibunit}

\end{document}